\documentclass[aps,twocolumn,superscriptaddress,prr,nofootinbib,longbibliography]{revtex4-2}
\usepackage{graphicx,graphics,xcolor,amsmath,mathtools,
subfigure,bm,bbm,tensor,bbold,braket,amssymb,xfrac}
\usepackage{hyperref}
\usepackage{cleveref}
\usepackage{stmaryrd}
\usepackage{dcolumn,kbordermatrix}
\usepackage{blkarray,float}

\newcommand{\doubleunderline}[1]{\underline{\underline #1}}

\crefname{equation}{Eq.}{Eqs.}
\Crefname{equation}{Equation}{Equations}
\crefname{figure}{Fig.}{Figs.}
\Crefname{figure}{Figure}{Figures}
\crefname{section}{Sec.}{Secs.}
\crefname{subsection}{Subsec.}{Subsecs.}
\Crefname{section}{Section}{Sections}
\crefname{appendix}{Appendix}{Appendices}
\Crefname{appendix}{Appendix}{Appendices}

\begin{document}

\title{ 
Operating a passive on-chip superconducting circulator: device control and quasiparticle effects}

\author{Dat Thanh Le} \email[]{thanhdat.le@uq.net.au}
\affiliation{ARC Centre for Engineered Quantum System, School of Mathematics and Physics,
University of Queensland, Brisbane, QLD 4072, Australia}
\author{Clemens M{\"u}ller}\email{Current address: Zurich Instruments AG, Zurich, Switzerland}
\affiliation{ARC Centre for Engineered Quantum System, School of Mathematics and Physics,
University of Queensland, Brisbane, QLD 4072, Australia}
\affiliation{IBM Quantum, IBM Research - Zurich, 8803 R{\"u}schlikon, Switzerland}
\author{Rohit Navarathna}
\affiliation{ARC Centre for Engineered Quantum System, School of Mathematics and Physics,
University of Queensland, Brisbane, QLD 4072, Australia}
\author{Arkady Fedorov}
\affiliation{ARC Centre for Engineered Quantum System, School of Mathematics and Physics,
University of Queensland, Brisbane, QLD 4072, Australia}
\author{T. M. Stace} \email[]{stace@physics.uq.edu.au}
\affiliation{ARC Centre for Engineered Quantum System, School of Mathematics and Physics,
University of Queensland, Brisbane, QLD 4072, Australia}

\begin{abstract}

Microwave circulators play an important role in quantum technology based on superconducting circuits. The conventional circulator design, which employs ferrite materials, is bulky and involves strong magnetic fields, rendering it  unsuitable for integration on superconducting chips. One promising design for an on-chip superconducting circulator is based on a passive Josephson-junction ring.
In this paper, we consider two operational issues for such a device: circuit tuning and the effects of quasiparticle tunneling.
We compute the scattering matrix using adiabatic elimination and derive the parameter constraints to achieve optimal circulation.  We then numerically optimize the circulator performance over the full set of external control parameters, including gate voltages and flux bias, to demonstrate that this multi-dimensional optimization converges quickly to find optimal working points. We also consider the possibility of quasiparticle tunneling in the circulator ring and how it affects signal circulation. Our results form the basis for practical operation of a passive on-chip superconducting circulator made from a ring of Josephson junctions.  
    
\end{abstract}

\maketitle

\section{Introduction}

Microwave circulators are widely used in experiments with superconducting circuits \cite{Gu17}. They break Lorentz reciprocity \cite{Deak12} and facilitate unidirectional signal propagation, thus protecting fragile quantum systems from noise and enabling   discrimination between input and output fields for quantum-limited amplification \cite{Chapman17}. Commercially available circulators are typically realized  using ferrite materials and the Faraday effect to induce non-reciprocity \cite{Pozar11}.  This approach necessitates device dimensions of the order of the microwave wavelength, which poses a practical difficulty for integrating circulators with chip-based superconducting circuits.  
Furthermore, the strong magnetic fields in conventional circulators are incompatible with sensitive superconducting devices. Hence, a great deal of effort has been devoted to implementation of ferrite-magnet-free circulators exploiting various physical mechanisms, such as the quantum Hall effect \cite{Viola14,Mahoney17}, interfering parametric processes \cite{Kamal11,Sliwa15,Lecocq17}, temporal modulation of couplings \cite{Peterson19,Fang12,Kamal17,Estep14,Kerckhoff15,Chapman17,Roushan17}, noncommutation between frequency conversion and delay \cite{Rosenthal17}, and reservoir engineering \cite{Metelmann15,Fang17}.

Recently, M{\"u}ller \textit{et al} \cite{Muller18} analyzed a proposal for a superconducting Josephson-junction-ring circulator whose working principle parallels that of conventional ferrite circulators.  The Josephson junction ring is promising for quantum simulation and potential applications that require non-reciprocity \cite{Koch10,Fatemi21,Perea21}, as it is compatible with on-chip superconducting circuits and works passively, i.e., does not require an external drive. The physics behind non-reciprocal signal circulation in this device is the Aharonov-Bohm effect \cite{Koch10,Muller18}.
This effect (and the signal circulation) is strongly dependent on the external charge and flux biases, the signal frequency, as well as fabrication imperfections of the device parameters.

Because optimal circulator performance requires precise tuning of the external parameters, we here address  two related operational issues: (i) tuning to the ideal working point in the multi-dimensional space of the control parameters, and 
(ii) the effect of quasiparticle induced fluctuations \cite{Matveev93,Joyez94} on  the circulator. Tuning the device will likely be necessary in all implementations and given the numerous independent control parameters, (i) may present an operational challenge. (ii) has not been touched upon  in Ref.\ \cite{Muller18} which only showed resilience of signal circulation against perturbations in external biases. Unlike these parameter perturbations,
tunneling of a quasiparticle into/out of a superconducting island shifts the charge bias on that island by one electron worth of charge \cite{Court08,Lutchyn05,Lutchyn06}, which detunes the circulator away from its optimal operating points and impairs the tuning procedure (i).
Understanding the effect of quasiparticles is a step towards mitigating their impact on the device operation.

Therefore, in this paper we first consider optimization of the superconducting circulator proposed in Ref.\ \cite{Muller18}, that is, we describe a protocol for tuning the device in the multi-dimensional parameter space to find optimal operating points. To do this, we employ the adiabatic elimination procedure to  extract semi-analytic expressions for  the scattering matrix elements in the SLH input-output formalism \cite{Josh17,Muller17}. This allows us to deduce quantitative conditions for optimal circulation. 
We also present numerical optimization results based on a full treatment of the multi-level scattering problem. 
 The numerics are found to be in excellent agreement with the semi-analytical predictions specifying optimal working points for the circulator.
 
Second,  we address the effect of quasiparticles on the circulator efficiency.
 We show that due to tunneling of quasiparticles between different pairs of superconducting islands the Josephson-ring circulator in Ref.\ \cite{Muller18} has four accessible charge-parity sectors. Given the same working conditions and parameters, these sectors circulate signals with different efficiencies. Stochastic jumps among the sectors caused by quasiparticle tunneling events then may result in unstable operation of the circulator device. To mitigate these fluctuations, we propose to employ quasiparticle-trapping techniques \cite{Joyez94,Court08,Martinis21,Aumentado04,Sun12,Kalashnikov20} to suppress  quasiparticle population.

The structure of this paper is as follows. In \cref{sec:CircuitAndSLH} we present the circuit design of the passive on-chip superconducting circulator along with the SLH formalism to numerically calculate the scattering matrix elements. Then in \cref{sec:AEPredictions} we derive the scattering matrix elements exploiting the adiabatic elimination technique and determine the conditions for optimal circulation, followed by numerical optimization in \cref{sec:Optimization}. Section \ref{sec:QPPoisoning} analyzes quasiparticle tunneling in the circulator system. The paper is concluded in \cref{sec:Conclusion}. Appendixes provide detailed calculations and additional information for the results in the main text.  

\section{Circuit design and SLH formalism} \label{sec:CircuitAndSLH}

\begin{figure}[th]
    \centering
    \includegraphics[scale=0.267]{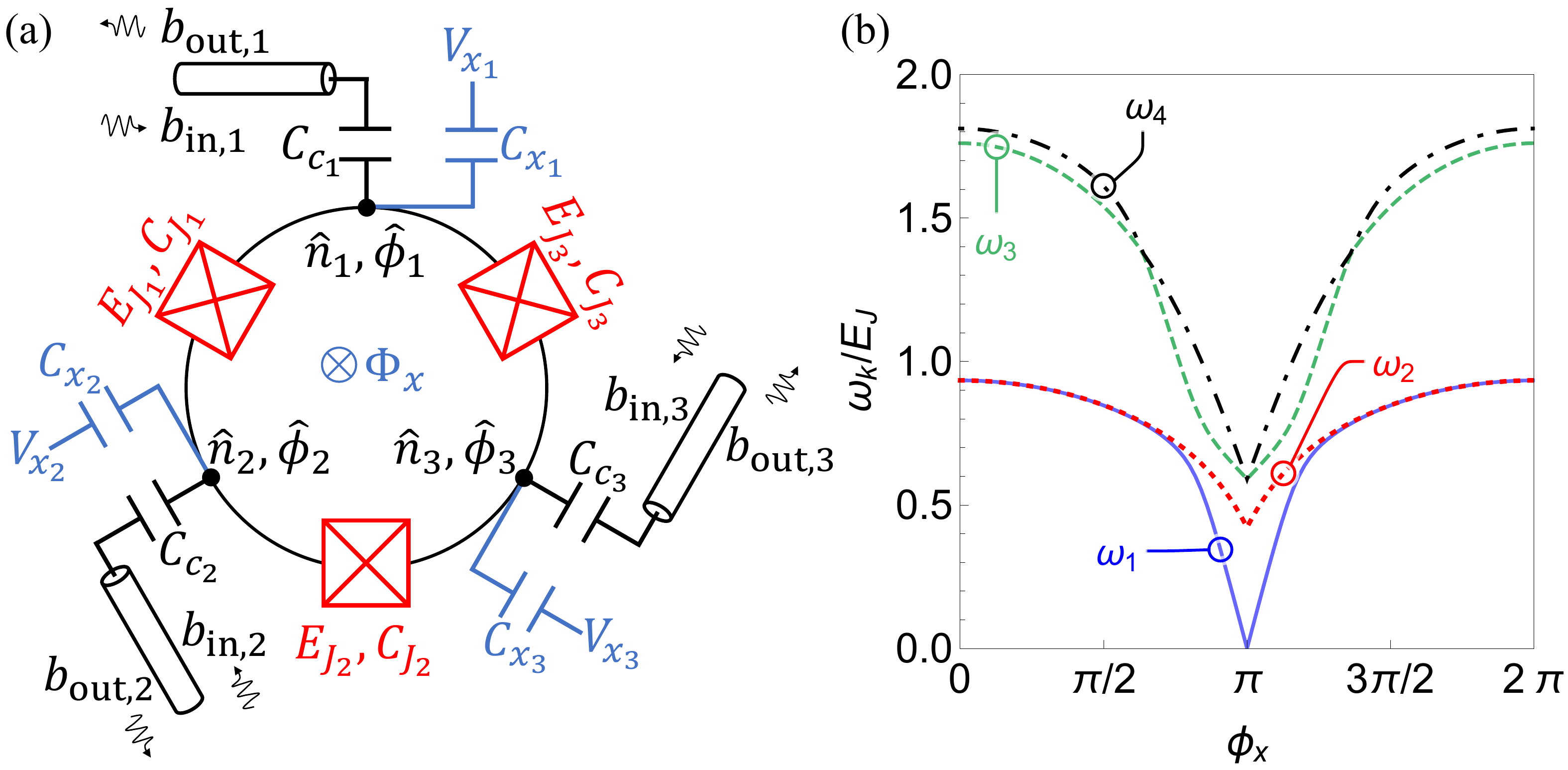}
    \caption{ (a) Schematic circuit design of the passive on-chip superconducting circulator proposed in Ref.\ \cite{Muller18}.  The device comprises  three superconducting islands which are represented by the numbers of Cooper pairs $\hat n_j$ and the superconducting phases $\hat \phi_j$ ($j=1,2,3$) on each island. They are connected by three Josephson junctions with Josephson energies $E_{J_j}$ and junction capacitances $ C_{J_j}$. Each island is biased by an external voltage $V_{x_j}$ via a gate capacitance $C_{x_j}$ and coupled capacitively to a waveguide via a coupling capacitance $C_{c_j}$. The whole circulator loop is threaded by a central external flux $\Phi_x$ as well.  (b)  First four excited-state energies $\omega_k$ ($k=1,2,3,4$) of the circulator ring versus the reduced external flux $\phi_x$ for a symmetric circuit (i.e., $E_{J_j} = E_J$, $C_{J_j}=C_J$, $C_{x_j} = C_x$, and $C_{c_j} = C_c$). 
    The eigenenergies are computed by numerically solving the eigensystem of $\hat H_{\mathrm{ring}}$ given in Eq.\ \eqref{eq:RingHamiltonian}  with $E_{C_\Sigma}/E_J = 0.35 $ and $n_{x_j} = 1/3$. 
    }
    \label{fig:circulator}
\end{figure}

In this section we present the circuit design of the ring circulator, its working principle, the SLH formalism to compute the scattering matrix, and the notations used throughout the paper. Many details of these can be found in Ref.\ \cite{Muller18}. The circulator circuit, depicted in \cref{fig:circulator}a, is a superconducting ring
segmented into three superconducting islands by three Josephson junctions each of which is described by a Josephson energy $E_{J_j}$ and a junction capacitance $C_{J_j}$ ($j=1,2,3$).  The three islands are represented by   the superconducting phases $\hat \phi_j$ and their conjugate charges $\hat n_j$; they are biased by external voltages $V_{x_j}$ with gate capacitances $C_{x_j}$ and coupled to three external waveguides by coupling capacitances $C_{c_j}$. The circulator ring is threaded by an external flux $\Phi_{x}$. Input fields $b_{\mathrm{in},j}$ propagate along the waveguides, interact with the ring, and scatter into output fields $b_{\mathrm{out},j}$. 

To begin, we consider the case of a symmetric Josephson-junction ring, that is, $E_{J_j} = E_J$ and $C_{J_j} = C_J$, and further assume that $C_{x_j} = C_x$ and $C_{c_j} = C_c$. We consider asymmetries later. As derived in \cref{append:CircuitQuantization}, the circulator ring Hamiltonian is 
\begin{eqnarray}
\hat H_{\mathrm{ring}} &=& \frac{(2e)^2}{2} (\hat{\mathbf{n}} - \mathbf{n}_x) \mathbb{C}^{-1} (\hat{\mathbf{n}} - \mathbf{n}_x) \nonumber \\
&& - E_J \sum_{j=1}^{3} \cos(\hat \phi_j - \hat \phi_{j+1} - \tfrac{1}{3} \phi_x),
\end{eqnarray}
where $\hat{\mathbf{n}} = \{ \hat n_1, \hat n_2, \hat n_3 \}$, $\mathbf{n}_x = \{ n_{x_1}, n_{x_2}, n_{x_3} \}$ with $n_{x_j} = C_{x_j} V_{x_j}/(2e)$ the (dimensionless) charge bias on the island $j$, $\phi_x = 2\pi \Phi_x/\Phi_0$ is the reduced flux bias which has been shared equally by the three Josephson junctions with $\Phi_0 = h/(2e)$ the superconducting quantum flux,  and $\mathbb{C}$ is the capacitance matrix. To account for the fact that the total number of Cooper pairs on the ring is conserved, we define new coordinates
\begin{eqnarray}
& \hat n'_1 = \hat n_1, \hspace{0.25cm} \hat n'_2 = - \hat n_2, \hspace{0.25cm} \hat n'_3 = \hat n_1 + \hat n_2 + \hat n_3 = n_0, \label{eq:CoorTransform} \\
& \hat \phi'_1 = \hat \phi_1 - \hat \phi_3, \hspace{0.25cm} \hat \phi'_2 = \hat \phi_3 - \hat \phi_2, \hspace{0.25cm} \hat \phi'_3 = \hat \phi_3, 
\end{eqnarray}
where $n_0$ is the conserved total charge number, which is controlled by the external biases \cite{Koch10}. In the new coordinates, the Hamiltonian $\hat H_{\mathrm{ring}}$ is 
\begin{eqnarray}
   \hat H_{\mathrm{ring}} &=& E_{C_{\Sigma}} \big( (\hat n'_1 - \tfrac{1}{2}(n_0 +n_{x_1}- n_{x_3}))^2 \nonumber  \\
&& + (\hat n'_2 + \tfrac{1}{2}(n_0 +n_{x_2}- n_{x_3}))^2 -\hat n'_1 \hat n'_2 \big) \nonumber \\
&& -E_J \big ( \cos(\hat \phi'_1 - \tfrac{1}{3}\phi_x) +\cos(\hat \phi'_2- \tfrac{1}{3}\phi_x) \nonumber \\
&& + \cos(\hat \phi'_1 + \hat \phi'_2  + \tfrac{1}{3}\phi_x) \big) \label{eq:RingHamiltonian},
\end{eqnarray}
where $E_{C_{\Sigma}} = (2e)^2/ C_{\Sigma}$ is the charging energy with \mbox{$C_{\Sigma} = 3C_J + C_x + C_c $}.

In terms of the ring eigenbasis $\{ \ket{k}; k=0,1,2, \dots \}$, we have
\begin{equation}
 \hat H_{\mathrm{ring}} =   \sum_{k>0} \omega_k \ket{k}\bra{k},
\end{equation}
 where $\omega_k$ is the eigenenergy\footnote{In this paper, we set $\hbar=1$.} associated with the excited state $\ket{k}$ ($k>0$), and we have subtracted the ground state energy of Eq.\ \eqref{eq:RingHamiltonian}, so that $\omega_0=0$.  Then $\omega_k$ represent ground to excited state transition frequencies that would be observed in spectra. In \cref{fig:circulator}b we plot the first four excited-state energies $\omega_k$ ($k=1,2,3,4$) as a function of the reduced external flux $\phi_x$.  These eigenenergies are arranged in pairs; for large ranges of $\phi_x$, $\omega_1$ and $\omega_2$ are nearly degenerate and so are $\omega_3$ and $\omega_4$. Circulation of signals in the device is mediated by these excitations: depending on the external biases and the driving frequency, signals emitted from different excitations interfere constructively/destructively resulting in clockwise/counter-clockwise circulation \cite{Viola14,Muller18}. This resembles the operation of a ferrite circulator where non-reciprocal transmission is created by interference of nearly degenerate, counter-propagating modes \cite{Pozar11}. 

To compute output fields scattering from the circulator, we make use of the SLH framework \cite{Josh17,Muller17}. We derive a Hamiltonian
description of  quantized bosonic fields for the waveguides interacting with the ring system.   The total Hamiltonian for the combined system is (see \cref{append:CircuitQuantization} for derivation)
\begin{equation}
    \hat H_{\mathrm{tot}} = \hat H_{\mathrm{ring}} + \hat H_{\mathrm{wg}} + \hat H_{\mathrm{int}}, \label{eq:TotHamiltonian}
\end{equation}
where $\hat H_{\mathrm{ring}}$ is given in Eq.\ \eqref{eq:RingHamiltonian} and the waveguide Hamiltonian $\hat H_{\mathrm{wg}}$ is 
\begin{equation}
\hat H_{\mathrm{wg}} = \sum_{j=1}^3 \int_{-\infty}^{\infty} d \omega \omega \hat a_j^\dag (\omega) \hat a_j (\omega), \label{eq:Hwg} \\
\end{equation}
which is the sum of three independent continua of harmonic oscillator modes. The interaction Hamiltonian $\hat H_{\mathrm{int}}$, under the Markov and rotating wave approximations, is \cite{Muller18,Gardiner85}
\begin{equation}
    \hat H_{\mathrm{int}} = \sum_{j=1}^3 \sqrt{\frac{\Gamma}{2\pi}} \int_{-\infty}^{\infty} d \omega (\hat a_j^\dag (\omega) \hat q_{j,-} + \hat a_j(\omega) \hat q_{j,+}  ), \label{eq:Hint}
\end{equation}
where $\hat q_{j,-}  \equiv (\hat q_{j,+})^\dag = \sum_{k<\ell} \langle k | \hat q_j | \ell \rangle \ket{k}\bra{\ell} $ is the upper triangularized part (in the ring eigenstate basis) of $\hat q_j$ \cite{Muller18} which is the coupling operator given in terms of  the charge operators as
 \begin{equation}
\begingroup 
\setlength\arraycolsep{1pt}
\begin{array}{ccc}
\hat  q_1 &=& \hat n'_1 + n'_{x_1} \\
&=& \hat n_1 + n'_{x_1}
\end{array}, \hspace{0.2cm} \begin{array}{ccc}
\hat q_2 &=& - \hat n_2' + n'_{x_2} \\
&=& \hat n_2 + n'_{x_2}
\end{array}, \hspace{0.2cm} \begin{array}{ccc}
    \hat q_3 &=& -\hat n'_1 + \hat n'_2 + n'_{x_3} \\
    &=& \hat n_3 - n_0+ n'_{x_3}
\end{array}.  
\endgroup \label{eq:CouplingOperator}
\end{equation}
 Here $n'_{x_j}$ are the rescaled charge biases
 \begin{equation}
     \left\{  \begin{array}{ccl}
         n'_{x_1} &=& c_1 (n_0 - n_{x_2} - n_{x_3}) - c_2 n_{x_1} \\
         n'_{x_2} &=& c_1 (n_0 - n_{x_1} - n_{x_3}) - c_2 n_{x_2} \\
         n'_{x_3} &=& c_2 (n_0 - n_{x_3}) - c_1 (n_{x_1} + n_{x_2})
    \end{array} \right. ,
 \end{equation}
 where $c_1 = C_J/(C_x + C_c)$, and $c_2 = (C_J +C_x + C_c)/(C_x + C_c)$. 
In \cref{eq:Hint}  $\Gamma$ is the waveguide-ring coupling strength explicitly given by \cite{Le19,Peropadre13,Devoret07,Blais21}
\begin{equation}
    \Gamma = 16 \frac{Z_{\mathrm{wg}}}{R_K} \left( \frac{C_c}{C_{\Sigma}} \right)^2  \omega_d = 32 \alpha \frac{Z_{\mathrm{wg}}}{Z_{\mathrm{vac}}} \left( \frac{C_c}{C_{\Sigma}} \right)^2 \omega_d ,  \label{eq:CouplingExpression}
\end{equation}
 where $Z_{\mathrm{wg}}$ is the waveguide impedance, \mbox{$R_K = h/e^2 \approx 25.8 \, \mathrm{k}\Omega$} is the resistance quantum,  \mbox{$\alpha = \mathrm{Z_{\mathrm{vac}}}/(2R_K) \approx 1/137$} is the fine-structure constant with $Z_{\mathrm{vac}} \approx 377 \,\Omega$ the vacuum impedance, and $\omega_d$ is the driving frequency.  
 As $C_c /C_{\Sigma} <1 $ by definition, for the typical situation of $Z_{\mathrm{wg}} = 50\, \Omega$ one finds $Z_{\mathrm{wg}}/Z_{\mathrm{vac}} \approx 0.13 $ and therefore $\Gamma <  0.03\, \omega_d $ justifying the approximations used to derive $\hat H_{\mathrm{int}}$. This holds for $Z_{\mathrm{wg}}\lesssim Z_{\mathrm{vac}}$ but may not for high-impedance waveguides \cite{Wiegand21}.
 The coupling strength $\Gamma$ additionally (as shown later)  sets the scale for resonance conditions and acceptable parameter imperfections in the circulator ring.

 Using the above Hamiltonians and considering single-mode weak coherent fields at the input ports with the amplitudes $\beta_j$ and the frequency $\omega_d$, the SLH master equation for the circulator density operator $\rho$ is given by \cite{Josh17,Muller17,Muller18}
\begin{equation}
    \dot \rho = - i [\hat H_{\mathrm{ring}} + \hat H_{\mathrm{drive}}, \rho ] + \sum_{j=1}^3 \mathcal D [\hat b_{\mathrm{out},j}] \rho, \label{eq:SLHme}
\end{equation}
where
\begin{eqnarray}
    \hat H_{\mathrm{drive}} &=& -\frac{i}{2} \sqrt{\Gamma} \sum_{j=1}^3 (\beta_j e^{-i \omega_{d} t } \hat q_{j,+} - \mathrm{H.c.} ), \label{eq:Hdrive}  \\
    \hat b_{\mathrm{out},j} &=&  \beta_j e^{-i \omega_{d} t } \mathbb{1} + \sqrt{\Gamma} \hat q_{j,-}, \label{eq:InputOutput}
\end{eqnarray}
and $\mathcal D[\hat {\mathcal O}] \rho = \tfrac{1}{2} (2 \hat{\mathcal O} \rho \hat{\mathcal O}^\dag - \rho \hat{\mathcal O}^\dag \hat{\mathcal O} - \hat{\mathcal O}^\dag \hat{\mathcal O} \rho ) $.
In \cref{eq:SLHme}, the commutation represents coherent evolution of the ring system plus the effect of dynamics induced from the external driving fields which is described by $\hat H_{\mathrm{drive}}$ in Eq.\ \eqref{eq:Hdrive}, whereas the dissipation is due to couplings to the waveguides. Equation \eqref{eq:InputOutput} represents the standard input-output relation \cite{Josh17,Gardiner85} in which the output field is the sum of the input field and the field radiated from the ring system.    

\section{Scattering matrix elements} \label{sec:AEPredictions}

We define the scattering matrix element $S_{ij}$ for transfer of signals from port $j$ to port $i$ as the ratio of the outgoing amplitude to the incoming one
\begin{equation}
S_{ij} =  \frac{ \langle \hat b_{\mathrm{out},i} \rangle  } { \langle \hat b_{\mathrm{in},j} \rangle } ,
\end{equation}
 where $\langle \hat {\mathcal{O}} \rangle = \mathrm{Tr} (\hat {\mathcal O} \rho) $ denotes the expectation value of an operator $\hat{\mathcal O}$ with $\rho$ the circulator ring density operator. $S_{ij}$ can be computed numerically by solving $\rho$ using the master equation in \cref{eq:SLHme}. However, 
in this section we harness the adiabatic elimination technique \cite{Josh17,Muller17}, which allows us to express scattering of the  open waveguide-ring system in terms of the isolated ring excitations, to derive a semi-analytical expression for $S_{ij}$. This expression precisely describes the working principle of the circulator and helps to find the
conditions to obtain optimal circulation.

\subsection{Adiabatic elimination}

 When a quantum system can be decomposed into a fast subspace $\mathcal F$ and a slow subspace $\mathcal S$, we can adiabatically eliminate its fast dynamics and consider its slow dynamics only \cite{Josh17}.
 For the circulator ring system, its fast subspace consists of the excited states $ \mathcal F = \{ \ket{k}, k>0 \}$, whereas its slow subspace contains the ground state only $\mathcal S = \{ \ket{0} \}$ \cite{Muller18}. In \cref{append:AE} we outline the calculations  for performing the adiabatic elimination on the circulator system. We find the scattering matrix element $S_{ij}$ restricted to the  slow subspace as
\begin{equation}
     S_{ij} =   \delta_{ij}  - \sum_{k>0} \frac{  \langle k | \hat q_j | 0\rangle \langle 0 | \hat q_i | k\rangle }{i\Delta \omega_k/\Gamma + \gamma_{k}/2} , \label{eq:Selement}
\end{equation}
where $ \langle k | \hat q_j |0 \rangle$ is the excitation amplitude due to the coupling operator $\hat q_j$, $ \langle 0 | \hat q_i | k\rangle$ is the relaxation amplitude 
due to the coupling operator $\hat q_i$, $\Delta \omega_k = \omega_k - \omega_d$ is the detuning of the excited eigenenergy $\omega_k$ from the driving frequency $\omega_d$, and $\gamma_{k} = \sum_{j=1}^3 |\langle 0 | \hat q_j | k \rangle |^2  $ represents the total (dimensionless) decay rate of the excited state $\ket{k}$ due to waveguide couplings. Similar expressions to \cref{eq:Selement} can be found in related works \cite{Koch10,Richman21} but for different circulator systems and using different derivation methods. The delta function $\delta_{ij}$ in \cref{eq:Selement} is a  consequence of the input-output relation in Eq.\ \eqref{eq:InputOutput}, in which the input field at one port contributes to the output field at that port, whereas the second term in \cref{eq:Selement} describes  interference via the transient excitations of the circulator ring. Equation \eqref{eq:Selement} demonstrates the importance of the external biases on signal scattering: they set the  values of the matrix elements $\langle k | \hat q_j | 0\rangle$ as well as the transition energy $\omega_k$ (and subsequently the detuning $\Delta \omega_k$). Therefore, precise control over these biases is  necessary to observe good circulation in the device.

At this point, it is instructional to consider the coherent power transmission of the scattered signals, \mbox{$P_j = \sum_{i=1}^{3} |S_{ij}|^2$}. Taking $|S_{ij}|^2$ in Eq.\ \eqref{eq:Selement} and summing over $i$, we find that 
\begin{eqnarray}
P_j = \sum_{i=1}^3 |S_{ij}|^2 &=& 1 - \sum_{k>0} \frac{ | \langle 0 | \hat q_j | k \rangle |^2     \gamma_{k}}{ \big( \tfrac{i}{\Gamma} \Delta \omega_k \!+\! \tfrac{1}{2} \gamma_{k}  ) ( - \tfrac{i}{\Gamma} \Delta \omega_k \!+\! \tfrac{1}{2} \gamma_{k}) } \nonumber \\
&& + \sum_{k,\ell>0} \frac{ \langle k | \hat q_j | 0 \rangle \langle 0 | \hat q_j | \ell \rangle Q_{k\ell} }{(\tfrac{i}{\Gamma} \Delta \omega_k \! + \! \tfrac{1}{2}\gamma_{k}) (- \tfrac{i}{\Gamma} \Delta \omega_{\ell} \!+\! \tfrac{1}{2}\gamma_{ \ell})}, \nonumber \\ \label{eq:SumColumn}
\end{eqnarray}
where $Q_{k \ell} = \sum_{j=1}^3 \langle 0 | \hat q_j | k \rangle \langle \ell | \hat q_j | 0 \rangle$ with $k,\ell>0$. Numerically, we observe that $|Q_{k\ne\ell}| \ll |Q_{kk}| \equiv \gamma_k$ (see \cref{fig:additionalInformation} in \cref{append:explainConditions}). 
Hence, in the second line of Eq.\ \eqref{eq:SumColumn} we can ignore terms with $k \ne  \ell$ and consider only those with $k = \ell$. We find that
\begin{equation}
   P_j = \sum_{i=1}^{3} |S_{ij}|^2 = 1, \label{eq:ColumnSum}
\end{equation}
which merely reflects the energy conservation constraint. We note that if incoherent scattering occurs (due to dephasing etc) then the coherent power transfer condition relaxes to  $P_j<1$, i.e.\  scattering into incoherent channels would appear as loss of total power in the coherent subspace.

\subsection{Conditions for optimal circulation} \label{subsec:Conditions}

\begin{figure}[t]
    \centering
    \includegraphics[width=0.48 \textwidth]{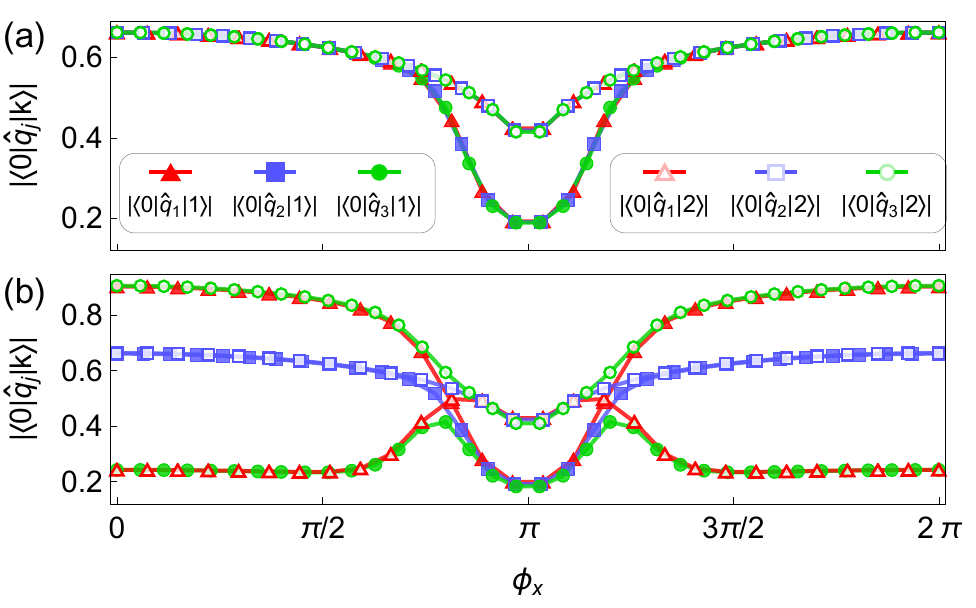}
    \caption{  Magnitudes of the coupling matrix elements $|\langle 0 | \hat q_j | 1 \rangle |$ (solid markers) and $|\langle 0 | \hat q_j | 2 \rangle |$ (open markers) with $j=1,2,3$ as functions of the reduced external flux $\phi_x$ when biasing the three ring islands identically at charge biases of $1/3$ for (a) a symmetric circulator ring with $E_{J_j}=E_J$ and $E_{C_{\Sigma}}/E_J = 0.35$ and for (b) an asymmetric circulator ring with $E_{J_1}/E_J=1$, $E_{J_2}/E_J = 1.01$, $E_{J_3}/E_J =0.99$, and $E_{C_\Sigma}/E_J = 0.35$.
    }
    \label{fig:CheckDipoleSymAsym}
\end{figure}

Based on the results in the previous subsection, we deduce the conditions for achieving optimal clockwise circulation. Note that the conditions for optimal counter-clockwise circulation can be found in a similar manner. We first introduce the scattering matrix for ideal (clockwise) circulation
\begin{equation}
 S_{\mathrm{ideal}}  = \left( \begin{array}{ccc}
0 & 1 & 0 \\
0 & 0 & 1 \\
1 &0 & 0
\end{array}
 \right), \label{eq:idealScatteringElement}
\end{equation}
noting that we are indifferent to the output phases of the non-zero elements. Since a diagonal element $S_{jj}$ from \cref{eq:Selement} is given by 
\begin{equation}
    S_{jj} =  1 - \sum_{k>0} \frac{ | \langle 0 | \hat q_j | k \rangle |^2 }{i\Delta \omega_k/\Gamma + \gamma_{k}/2}, \nonumber 
\end{equation}
to have $S_{11}=S_{22}=S_{33}$ one needs
\begin{equation}
| \langle 0 | \hat q_1 | k \rangle | =  | \langle 0 | \hat q_2 | k \rangle |=  | \langle 0 | \hat q_3 | k \rangle | \hspace{0.5cm} \mathrm{for\,} k>0. \label{eq:dipoleCondition}
\end{equation}
From \cref{eq:CouplingOperator} we have $ |\langle 0 | \hat q_j | k \rangle | = | \langle 0 | \hat n_j | k \rangle | $ with $\hat n_j$ the original charge operator on the island $j$, so the above condition is equivalent to $| \langle 0 | \hat n_1 | k \rangle | =  | \langle 0 | \hat n_2 | k \rangle |=  | \langle 0 | \hat n_3 | k \rangle |$ suggesting that the three  islands of the circulator ring should be symmetric.\footnote{When the islands of the ring circulator are symmetric, its Hamiltonian is invariant with respect to cyclic permutations of the node labels, $j=1\to 2 \to 3 \to 1$ or $j=1 \to 3 \to 2 \to 1$. Under these permutations, $\hat n_j$ becomes $\hat n_{j'}$, the ground state $\ket{0}$ is unchanged, and the excited state $\ket{k}$ picks up a phase, which results in $| \langle 0 | \hat n_j | k \rangle | = | \langle 0 | \hat n_{j'} | k \rangle |$.    } In the case of a symmetric circulator ring with identical Josephson junctions, this implies that the charge biases on the islands should also be identical.
In \cref{fig:CheckDipoleSymAsym} we plot $| \langle 0 | \hat q_j | 1 \rangle | $ (solid markers) and $| \langle 0 | \hat q_j | 2 \rangle | $ (open markers)  with $j=1,2,3$ versus the reduced external flux $\phi_x$ for both symmetric (panel a) and asymmetric (panel b) circulator rings at identical charge biases $n_{x_j}$ of $1/3$.
 For a symmetric ring, we observe in \cref{fig:CheckDipoleSymAsym}a that the condition in  \cref{eq:dipoleCondition}  is  satisfied  for the whole range of $\phi_x$ from zero to $2\pi$. 
 For an asymmetric ring with different Josephson energies in \cref{fig:CheckDipoleSymAsym}b, the condition in \cref{eq:dipoleCondition} is approximately met for a small interval around $\phi_x = \pi$.

From \cref{eq:CouplingOperator}, we have \mbox{$\hat q_3 = - \hat q_1 - \hat q_2 + \sum_{j=1}^3 n'_{x_j}  $}.
By this we can recast the condition in \cref{eq:dipoleCondition} to \mbox{$ | \langle 0 | \hat q_1 | k \rangle | = | \langle 0 | \hat q_2 | k \rangle  | = | \langle 0 | \hat q_1 | k \rangle + \langle 0 | \hat q_2 | k \rangle |$}. 
This is then translated into the following conditions
\begin{eqnarray}
\langle 0 | \hat q_j | k \rangle &=& r_{k} e^{i \varphi_{j,k}} \hspace{0.67cm} \mathrm{for\,} j=1,2,  \label{eq:nCondition} \\
| \varphi_{1,k} - \varphi_{2,k} | &=& \frac{2\pi}{3}  \hspace{1.33cm} \mathrm{for\,} k>0. \label{eq:phaseCondition}
\end{eqnarray}
The former condition tells us that the matrix elements between the ground state $\ket{0}$ and the excitation $\ket{k}$ of the coupling operators $\hat q_1$ and $\hat q_2$ should have the same magnitude, while the latter imposes a specific constraint on the phases of these matrix elements.
These two conditions were pointed out in Ref.\ \cite{Koch10} based on a system that includes external cavities on the output of each circulator waveguide, whereas our analysis is based directly  on the circulated scattering elements.

We next derive
the relations between the driving frequency $\omega_d$, the coupling strength $\Gamma$, and the first two transition energies $\omega_1$ and $\omega_2$ to observe optimal circulation. 
We notice that the strong anharmonicity of the circulator ring (see \cref{fig:circulator}b) allows us to consider contributions of only the first two excitations to signal circulation and ignore those of higher excitations; thus, in \cref{eq:Selement} the values of $k$ are truncated to $\{1,2 \}$.  We define new parameters
\begin{equation}
    x_k =  \frac{|r_k|^2}{((\Delta\omega_k/\Gamma)^2 + (\gamma_{k}/2)^2)^{1/2}}, \hspace{0.3cm} \tan (\theta_k) = \frac{-2 \Delta\omega_k}{\Gamma \gamma_{k}}, \label{eq:DefNewPara}
\end{equation}
for $k=1,2$. From Eqs.\ \eqref{eq:nCondition} and \eqref{eq:phaseCondition} we recast $S_{ij}$ in terms of $r_k$ and $\varphi_{j,k}$ and subsequently $x_k$ and $\theta_k$. For example,
we find \mbox{$S_{11} = 1 - x_1 e^{i\theta_1} - x_2 e^{i\theta_2} $} and \mbox{$S_{21} = x_1 e^{i(\theta_1 \pm_1 2\pi/3)} + x_2 e^{i(\theta_2 \pm_2 2\pi/3)}$}, where the signs $\pm_{k}$ can be different between the levels \cite{Koch10}.  Using \cref{eq:ColumnSum}, 
 the first column of the ideal scattering matrix $(S_{11},S_{21},S_{31}) = (0,0,1)$ is equivalent to \mbox{$(S_{11},S_{21}) = (0,0)$}, yielding
\begin{eqnarray}
  S_{11} &=& 1-   x_1 e^{i\theta_1} - x_2 e^{i\theta_2} = 0, \\
    S_{21} &=&   x_1 e^{i(\theta_1 - 2\pi/3)} + x_2 e^{i(\theta_2 + 2\pi/3)} = 0,
\end{eqnarray}
where we have chosen specifically the sign of $\pm_k$ in the phase factors of $S_{21}$. 
  The solution for this system of equations is
\begin{equation}
 x_1 = x_2 =1/ \sqrt{3}, \hspace{0.9cm}
    \theta_1 = -\theta_2 = \pi/6,  \label{eq:Solution}
\end{equation}
which results in
\begin{equation}
    -\Delta \omega_1 = \frac{1}{2\sqrt{3}} \gamma_1 \Gamma, \hspace{0.9cm} \Delta \omega_2 = \frac{1}{2\sqrt{3}} \gamma_2 \Gamma. \label{eq:Consequence}
\end{equation}
We recall that $\gamma_1$ and $\gamma_2$ respectively represent the decay rates of the first two excited states $\ket{1}$ and $\ket{2}$. We aim to operate the circulator at the parameter ranges such that the two excited states are nearly degenerate, so we can have $\gamma_1 \simeq \gamma_2 = \gamma$.
This, combining with the results in \cref{eq:Consequence},  yields
\begin{eqnarray}
\omega_d &\simeq& \frac{1}{2}(\omega_1 + \omega_2), \label{eq:driveCondition} \\
 \Gamma &\simeq& \frac{\sqrt{3}}{\gamma}  (\omega_2 - \omega_1). \label{eq:couplingCondition}
\end{eqnarray}
The former condition
ensures that the driving fields excite the first two nearly-degenerate excited states equally. Meanwhile, the latter introduces a concrete relation between the coupling strength $\Gamma$ and the eigenenergy difference $\omega_2 - \omega_1$ \cite{Richman21}, which can be met by suitably tuning the reduced external flux $\phi_x$.
Note that the same results in Eqs.\ \eqref{eq:Solution} to \eqref{eq:couplingCondition} are obtained when using either the second column or the third column of $S_{\mathrm{ideal}}$.

Based on the above conditions, we implement a simple numerical scheme to compute the optimal parameters for circulation. Considering a symmetric Josephson-junction ring with identical Josephson energies, the condition in Eq.\ \eqref{eq:dipoleCondition} indicates that we should choose identical charge biases (for example, at $1/3$ of a Cooper pair), while the driving frequency $\omega_d$ should be chosen to be $(\omega_1+\omega_2)/2$ as suggested by the condition in Eq.\ \eqref{eq:driveCondition}. 
 The external flux $\phi_x$ is determined via the condition in Eq.\ \eqref{eq:couplingCondition}. Noting that the charge offsets are already fixed ($n_{x_j} = 1/3$), the transitions $\omega_1$ and $ \omega_2$, the decay rate $ \gamma,$ and the coupling $\Gamma$ are implicitly functions of $\phi_x$. Then the optimal value for $\phi_x$ is numerically found from the equation
 $\Gamma (\phi_x) = \sqrt{3} (\omega_2 (\phi_x) - \omega_1 (\phi_x))/\gamma(\phi_x)$. For an asymmetric ring with different Josephson energies, it is no longer straightforward to estimate the optimal charge biases analytically. However we can consider the relevant quantities  as functions of the charge biases $n_{x_j}$ and the external flux $\phi_x$.\footnote{As $\Gamma$ is given in terms of $\omega_d$ as in \cref{eq:CouplingExpression} and $\omega_d$ is chosen to be $(\omega_1+\omega_2)/2$ depending on $\phi_x$, $\Gamma$ is treated as a function of $\phi_x$. } We evaluate the optimal working point by numerically finding $n_{x_j}$ and $\phi_x$ that satisfy the conditions in Eqs.\ \eqref{eq:dipoleCondition}, \eqref{eq:driveCondition}, and \eqref{eq:couplingCondition}.

\section{Optimization of operating parameters} \label{sec:Optimization}

 We note that solving the conditions in Eqs.\ \eqref{eq:dipoleCondition}, \eqref{eq:driveCondition}, and \eqref{eq:couplingCondition} gives physical insights into the optimal working parameters above. However extracting the quantities such as $\omega_1$, $\omega_2$, and $\Gamma$ from experiments to sufficiently high accuracy may be difficult in practice.  Therefore we now implement an  optimization procedure that finds the optimal working points using a standard optimization method.  We have checked that this approach gives the same result for $\phi_x$, $n_{x_j}$ and $\omega_d$ as solving Eqs.\ \eqref{eq:dipoleCondition}, \eqref{eq:driveCondition}, and \eqref{eq:couplingCondition}, as described in the previous section.

We optimize a cost function that  finds points of high fidelity $F(|S|,S_{\mathrm{ideal}})$ \cite{Scheucher16} between the computed scattering matrix $S$ and the ideal clockwise scattering matrix, $S_{\mathrm{ideal}}$, in \cref{eq:idealScatteringElement}.  We present the  optimization results for both symmetric and asymmetric circulator rings. 

We define the fidelity $F(A,B)$ between two matrices $A$ and $B$ as
\begin{equation}
    F(A,B) = 1- \frac{1}{ \lVert A \rVert \lVert B \rVert} \Big( \sum_{i,j} |A(i,j) - B(i,j)|^2 \Big)^{1/2}, \label{eq:DefFidelity}
\end{equation}
where $ \lVert X \rVert = \sqrt{ \mathrm{Tr}(X X^\dag)} $ denotes the norm of a matrix $X$. In \cref{eq:DefFidelity} 
the second term describes a distance measure between two matrices. The fidelity is thus  complementary to the distance measure: if two matrices are very similar to each other, their distance measure will be close to zero but their fidelity will be close to one.  It is worth noting that a non-circulating device with $S \simeq \mathbb{1}$ and a counter-clockwise-circulating one with $S \simeq  S_{\mathrm{ideal}}^\intercal $ have \mbox{$F(\mathbb{1}, S_{\mathrm{ideal}}) = F(S_{\mathrm{ideal}}^\intercal, S_{\mathrm{ideal}})  \approx 0.18$}, which sets the \textit{neutral} value of the fidelity.

We fix the energy scales of the circulator (i.e., the Josephson energies $E_{J_j}$ and the charging energy $E_{C_{\Sigma}}$) and employ a standard optimization routine (\textsf{FindMaximum} in \textsf{MATHEMATICA}) to optimize the fidelity over five control parameters, namely, the driving frequency $\omega_d$, the reduced external flux $\phi_{x}$, and the three charge biases $n_{x_1}$, $n_{x_2}$, and $n_{x_3}$. Over the course of optimization we also track the variations of other quantities, including the coupling matrix elements $\langle 0 | \hat q_j | k \rangle$, the coupling strength $\Gamma$, and the ring eigenenergies $\omega_1$ and $\omega_2$. We show that the optimization converges relatively fast after less than 50 optimization steps. In a realistic experiment, this would require sequential measurements of the full scattering matrix. Given typical experimental  time per single scattering matrix measurement of $10-100\, \mu\mathrm{s}$ \cite{Jerger16}, the total optimization would take $0.5 - 5\, \mathrm{ms}$, indicating feasibly fast calibration of the device.

\subsection{Symmetric Josephson-junction ring} \label{subsec:SymNum}

\begin{figure}[h!]
\centering 
\includegraphics[scale=0.82]{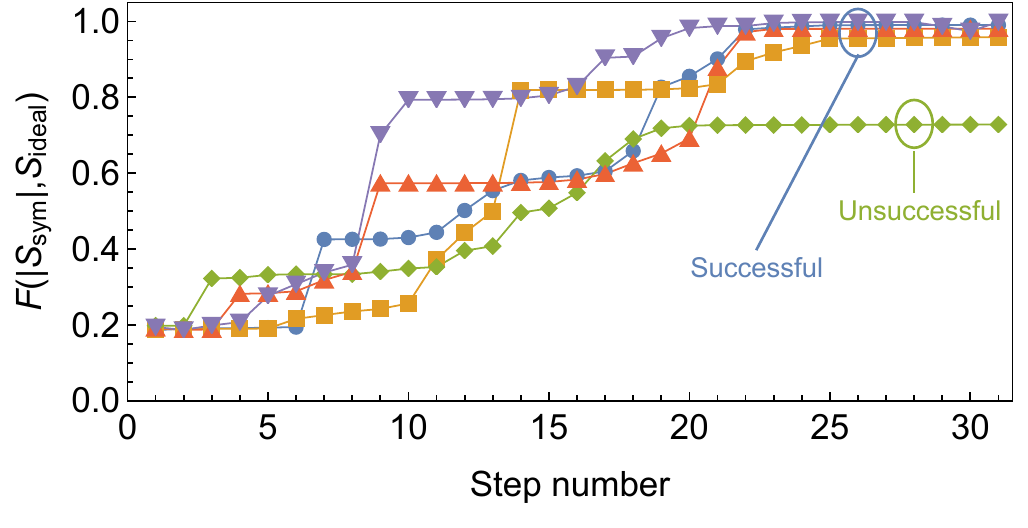}
 \caption{Examples of optimization for a symmetric circulator ring. We optimize the fidelity $F(|S_{\mathrm{sym}}|, S_{\mathrm{ideal}} )$ between the numerically computed scattering matrix $S_{\mathrm{sym}}$ with respect to the ideal (clockwise) one $S_{\mathrm{ideal}}$ for five times. 
 Each optimization begins with a different set of initial external parameters chosen randomly from specific ranges (see also main text) and takes 30 steps to complete. 
Relevant parameters are chosen as $E_{Jj} = E_J$, $E_{C_\Sigma}/E_J = 0.35$, and $\Gamma/E_J \approx 0.0025 $ for $Z_{\mathrm{wg}} = 50 \Omega$ and  $\omega_d/E_J = 0.8$. 
    }
   \label{fig:OptFideSym}
\end{figure}

We consider a symmetric Josephson-junction ring with $E_{J_j} = E_{J}$ ($j=1,2,3$), $E_{C_{\Sigma}}/E_J = 0.35$, and $\Gamma/E_J \approx 0.0025 $ for $Z_{\mathrm{wg}} = 50\, \Omega$ and  $\omega_d/E_J = 0.8$ and perform five optimizations for the fidelity $F(|S_{\mathrm{sym}}|,S_{\mathrm{ideal}})$. In \cref{append:SimuPara} we show specific parameter values for simulations.
Here the ratio $E_{C_\Sigma}/E_J = 0.35$ is chosen to be in-between the `Cooper-pair-box' ($E_{C_\Sigma}/E_J \gg 1$) and `transmon' ($E_{C_\Sigma}/E_J \ll 1$) regimes  for the following reasons. First,  operating the device outside the 'Cooper-pair-box' regime avoids charge sensitivity. Second, as pointed out in Ref.\ \cite{Koch10} reducing $E_{C_\Sigma}/E_J$  into the 'transmon' regime, which intuitively should make the  device insensitive to charge noise, actually destroys the circulation feature. This is because in this regime all the coupling matrix elements can be chosen to be purely imaginary breaking down the interference effect [see \cref{eq:Selement}]. Third, we find that when decreasing $E_{C_{\Sigma}}/E_J$  the decay rate $\gamma$ in Eq.\ \eqref{eq:couplingCondition} increases while the transition difference $\omega_2 - \omega_1$  decreases. Thus, reducing $E_{C_\Sigma}/E_J$ results in a small optimal coupling strength $\Gamma$ [which is proportional to $ (\omega_2 - \omega_1)/\gamma$ as in Eq.\ \eqref{eq:couplingCondition}] as well as a small working bandwidth. We confirm these numerically  in \cref{append:SimuPara}. Additionally, as shown later a small $\Gamma$  will put hard constraints on junction fabrication.

Each of the optimizations is initialized with a set of external parameters chosen randomly within certain ranges. That is $\omega_d/E_J \in [0.70,0.85]$, $\phi_x \in [1.00,2.14]$, and $n_{x_j} \in [0,1]$, reflecting the experimental uncertainties in initial parameters immediately after cooldown of the device, e.g., due to charge frozen in the substrate materials, flux defects, and charge-reset noise $\delta Q \sim \sqrt{k_BTC}$ worth approximately one electron for $\mathrm{fF}$ gate capacitors at the cooling temperature $T \sim 1 \mathrm{K}$ \cite{Johnson28,Nyquist28}. The ranges of $\omega_d$ and $\phi_x$ are intentionally selected such that $\omega_2 - \omega_1$ is neither too large nor too small compared to $\Gamma$, as suggested from Eq.\ \eqref{eq:couplingCondition} and observed from the circulator spectrum in \cref{fig:circulator}b. 
 We track the fidelity  during optimization steps to see how quickly the optimization proceeds. 
We also plot the optimization process for a representative selection of randomly initialized external control parameters in \cref{append:Inspect}.

 As shown in \cref{fig:OptFideSym} four out of five example optimizations yield a very high fidelity ($ \approx 1$), after 25 to 30 optimization steps.  In these cases the driving frequency $\omega_d$ and the reduced flux $\phi_x$ in Figs.\  \ref{fig:OptParaSym}a and \ref{fig:OptParaSym}b in \cref{append:Inspect}  evolve to well-defined values at about $ 0.82 \,E_J$ and $ 1.77$, respectively. Meanwhile, the three charge biases $n_{x_j}$ in Figs.\ \ref{fig:OptParaSym}c - \ref{fig:OptParaSym}e tend towards the same value with two apparent clusters near 0.4.
 
 The resulting power transfer matrix after the successful optimizations is 
\begin{equation}
\left| S_{\mathrm{opt}} \right|^2   \approx  \left( \begin{array}{ccc}
0.003 & 0.995 & 0.002 \\
0.002 &  0.003 & 0.995 \\
0.995 & 0.002 & 0.003
\end{array}
 \right). \label{eq:symScatMatrix}
\end{equation}
This corresponds to  insertion loss of  $\mathrm{IL}\approx 0.02$ dB while the reflection  and the isolation  are  $\mathrm{R}\approx-25$ dB and $\mathrm{IS}\approx -27$ dB respectively, where 
\begin{eqnarray}
\mathrm{IL} &=& 10\, \mathrm{Log}_{10} \big((|S_{12}|^2 + |S_{23}|^2 + |S_{31}|^2)/3\big), \\
\mathrm{R} &=& 10\, \mathrm{Log}_{10} 
\big( (|S_{11}|^2 + |S_{22}|^2 + |S_{33}|^2)/3\big), \\
\mathrm{IS} &=& 10\, \mathrm{Log}_{10} \big((|S_{13}|^2 + |S_{21}|^2 + |S_{32}|^2)/3\big).
\end{eqnarray}

In some cases the optimization can be trapped in a sub-optimal configuration; for example, the unsuccessful optimization (solid green diamond) in \cref{fig:OptFideSym} yields a substantially reduced fidelity ($ \approx 0.7$). In this scenario the three charge biases, as shown in Figs.\ \ref{fig:OptParaSym}c - \ref{fig:OptParaSym}e,  arrive at rather different final values,  partly explaining why the fidelity for that optimization is not as high as for the other optimizations. This failure, possibly, is due to our use of a very
simple optimization algorithm and may be circumvented
by repeating the optimization from a different starting parameter set or by employing more sophisticated parameter optimization
routines.
 
Furthermore,
we observe that after optimization the condition in Eq.\ \eqref{eq:dipoleCondition} is typically satisfied. This is demonstrated in Figs.\  \ref{fig:OptSymCheck}a - \ref{fig:OptSymCheck}f in   \cref{append:Inspect} which show that the matrix element magnitudes $|\langle 0 | \hat q_j | k \rangle|$ for $j=1,2,3$ and $k=1,2$  approach the same value for the successful optimizations.
We also confirm the conditions in Eqs.\  \eqref{eq:driveCondition} and \eqref{eq:couplingCondition} by plotting the two ratios  $2\omega_d/(\omega_1 + \omega_2)$ and $ \gamma \Gamma/(\omega_2 - \omega_1)$ respectively in  Figs.\ \ref{fig:OptSymCheck}g and \ref{fig:OptSymCheck}h.

\subsection{Asymmetric Josephson-junction ring} \label{subsec:AsymNum}

\begin{figure}[h!]
 \centering
    \includegraphics[scale=0.82]{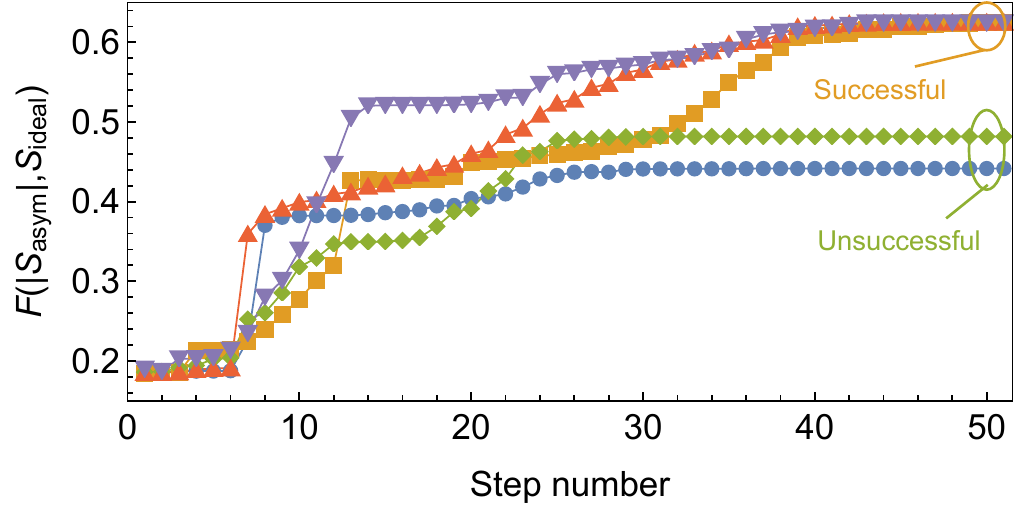}
      \caption{Similar to \cref{fig:OptFideSym} but for an asymmetric circulator ring. 
      Relevant parameters are chosen as $E_{J_1}/E_J=1$, $E_{J_2}/E_J = 1.01$, $E_{J_3}/E_J =0.99$, $E_{C_\Sigma}/E_J = 0.35$, and $\Gamma/E_J \approx 0.0025 $ for $Z_{\mathrm{wg}} = 50 \Omega$ and  $\omega_d/E_J = 0.8$.
    }
   \label{fig:OptFideASym}

\end{figure}

 Realistic device fabrication is always imperfect giving rise to junction asymmetry. Therefore, we introduce asymmetry in the circulator junctions as $E_{J_1} = E_{J}$, $E_{J_2} =E_J + \delta E_{J_2}$, and $E_{J_3} = E_J + \delta E_{J_3}$. To illustrate the effect of imperfect Josephson junctions we choose $\delta E_{J_2}/E_J = 0.01 $ and $\delta E_{J_3}/E_J = -0.01$ and  show the corresponding optimization results in \cref{fig:OptFideASym}.
Such junction asymmetry is plausible in realistic experiments \cite{Fink09,Osman21}. For example, Ref.\ \cite{Osman21} reported fabrication of on-chip Josephson junctions with high reproducibility and normal resistance ($R_N$) variation as small as $1.2\,\%$ which corresponds to $|\delta E_J|/E_J \sim  0.012$.\footnote{From the Ambegaokar–Baratoff relation \cite{Osman21} $E_J = A (\bar{R}_N + \delta R_N)^{-1}  = \bar{E}_J + \delta E_J $, where $A$ is a constant and $\bar E_J = A \bar R_N^{-1}$, we can estimate $|\delta E_J|/\bar E_J = |\delta R_N|/\bar R_N$. }  

In \cref{fig:OptFideASym} three out of five optimizations converge to fidelities just above $ 0.6$ after 50 steps of optimization.
The  power transfer matrix for these optimizations is
\begin{equation}
\left |S_{\mathrm{opt}} \right|^2  \approx  \left( \begin{array}{ccc}
0.08 & 0.70 & 0.22 \\
0.35 & 0.06 & 0.59 \\
0.57 & 0.24 & 0.19
\end{array}
 \right), 
\end{equation}
showing that the device circulates imperfectly with \mbox{$\mathrm{IL} \approx -2.1 \, \mathrm{dB} $}, $\mathrm{R} \approx -9.5\, \mathrm{dB} $, and $\mathrm{IS} \approx -5.7\, \mathrm{dB}$.

Similar to the symmetric case, the driving frequency $\omega_d$ and the reduced flux $\phi_x$ in Figs.\ \ref{fig:OptParaASym}a and \ref{fig:OptParaASym}b of \cref{append:Inspect} approach well-defined values at about $ 0.70 E_J$ and $ 2.41$, respectively. In contrast to the symmetric case, the three charge biases in Figs.\ \ref{fig:OptParaASym}c - \ref{fig:OptParaASym}e tend towards different values during optimizations. 
However, as shown in \cref{fig:OptASymCheck}  the conditions in Eqs.\ \eqref{eq:dipoleCondition} and \eqref{eq:driveCondition} are still approximately fulfilled: the matrix element magnitudes $|\langle 0 | \hat q_j | k \rangle|$ ($j=1,2,3$ and $k=1,2$) are quite close to each other (see Figs.\  \ref{fig:OptASymCheck}a - \ref{fig:OptASymCheck}f) and the ratio $2\omega_d/(\omega_1 + \omega_2)$ gets to almost exactly 1 (see \cref{fig:OptASymCheck}g). The ratio $\gamma \Gamma/(\omega_2 - \omega_1)$ in \cref{fig:OptASymCheck}h approaches about $0.45$, far below the optimal value $\sqrt{3}$ required in the condition in Eq.\ \eqref{eq:couplingCondition}.

\begin{figure}[ht!]
    \centering
    \includegraphics[scale=0.97]{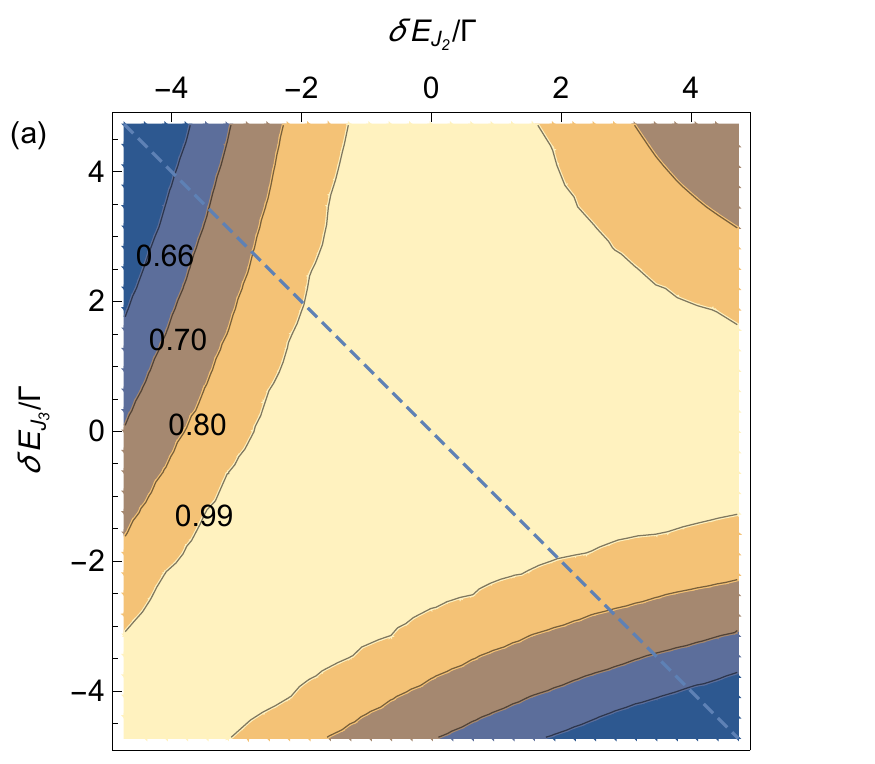}
    \includegraphics[scale=0.97]{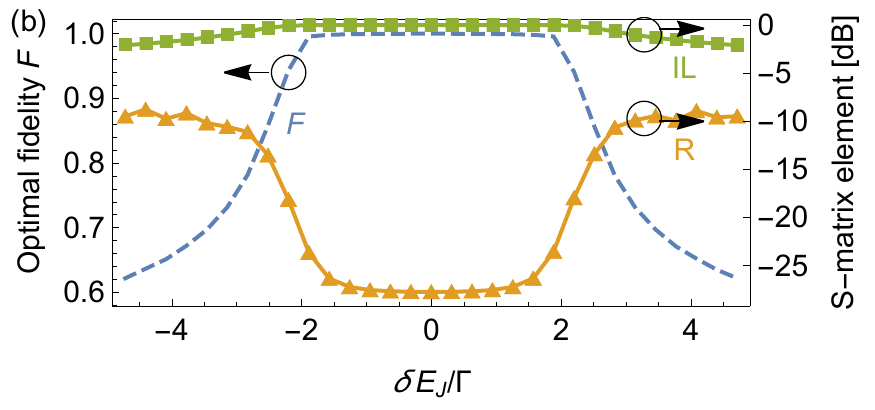}
    \caption{(a) Contour plot of the optimal fidelity, $F$, versus the ratios of the junction disorders $\delta E_{J_2}$ and $\delta E_{J_3}$ to the coupling strength $\Gamma$. (b) Diagonal cut of the optimal fidelity in panel (a) (blue dashed line) which corresponds to junction asymmetries $\delta E_{J_2} = -\delta E_{J_3} = \delta E_J$ considered in \cref{fig:OptFideASym}. Also shown in panel (b) includes  reflection, $\mathrm{R}$, and insertion loss, $\mathrm{IL}$. 
    To generate the plots, we introduce junction asymmetry as $E_{J_1} = E_J$, \mbox{$E_{J_2} = E_{J} + \delta E_{J_2}$}, and $E_{J_{3}} = E_{J} + \delta E_{J_3}$ and for each pair  $(\delta E_{J_2}, \delta E_{J_3})$ we find the optimal fidelity by performing optimization over the external control parameters and calculate the scattering matrix elements at the optimized working points. Here at the leftmost or rightmost of panel (b), $|\delta E_J|/\Gamma \approx 4.7 $ and the optimal fidelity is   $F\approx0.6$, consistent with the optimized value in \cref{fig:OptFideASym}.  }
    \label{fig:checkASYM}
\end{figure}

Comparing the fidelities in Figs.\ \ref{fig:OptFideSym} and \ref{fig:OptFideASym}, we observe a $40\%$ reduction in the optimized fidelity as a result of only $1\%$ asymmetry in the ring junctions. This follows from the fact that in this asymmetric case $\Gamma$ ($\approx 0.0025 E_J$) is substantially smaller than the detuning $\omega_1 - \omega_2$ ($\approx 0.01 E_J$) between the two excited states.  Thus, there is no driving frequency that simultaneously couples strongly to both states $\ket{1}$ and $\ket{2}$, and subsequently the condition for interference between these states is inhibited.  To show that $\Gamma$ sets the tolerance level for asymmetries in junction parameters, in \cref{fig:checkASYM}a we plot the optimal fidelity versus the two ratios  $\delta E_{J_2}/\Gamma$ and $\delta E_{J_3}/\Gamma$ and in \cref{fig:checkASYM}b we consider the example shown in \cref{fig:OptFideASym} with $\delta E_{J_2} = - \delta E_{J_3} = \delta E_{J}$. 
We see in Figs.\ \ref{fig:checkASYM}a and \ref{fig:checkASYM}b that the optimal fidelity remains close to 1 for $|\delta E_J|/\Gamma $ as large as $2$ but decreases quite quickly for larger $|\delta E_J|/\Gamma$. Accordingly, in \cref{fig:checkASYM}b the reflection is below $-20$ dB and the insertion loss is very close to $0$ dB for that range of $|\delta E_J|$. The parameters used for the optimizations in \cref{fig:OptFideASym} give $\delta E_J/\Gamma \approx 4.7$ corresponding to an optimal fidelity slightly above $0.6$, which is consistent with the values at the leftmost or rightmost of \cref{fig:checkASYM}b.  

\section{Quasiparticles} \label{sec:QPPoisoning}

In the previous sections, we have specified values for the external control parameters that optimize the circulator performance.
These parameters are subject to fluctuations due to voltage noise of various sources. Fast charge fluctuations with magnitudes much smaller than one electron have been studied in Ref.\ \cite{Muller18}, while charge drifts comparable to one electron are expected to occur at a timescale much longer than the optimization time.  Hence, in what follows we focus on
quasiparticle formation and migration in the superconducting islands forming the circulator, which is a frequently encountered noise source in superconducting devices \cite{Wilen21}.  We analyze the effects of quasiparticles on the circulator performance. We anticipate  quasiparticle formation will be slow relative to the internal dynamical timescales of the circulator, so our analysis is quasi-static.
Quasiparticle tunneling  causes \textit{large} changes in the effective bias voltages, on the scale of half a Cooper pair, and thus is non-perturbative.
We show that quasiparticle tunneling in the ring circulator results in several operating sectors characterized by the parity of charges on the islands. These sectors have different energy spectra and scatter signals differently,  yielding different circulation performances. Spectroscopic measurements performed on the circulator ring over experimental timescales much larger than the quasiparticle formation and tunneling rates will show a mixture of the spectra from the different quasiparticle sectors.

\subsection{Parity-charge sectors}

\begin{figure}[ht!]
    \centering
    \includegraphics[scale=0.393]{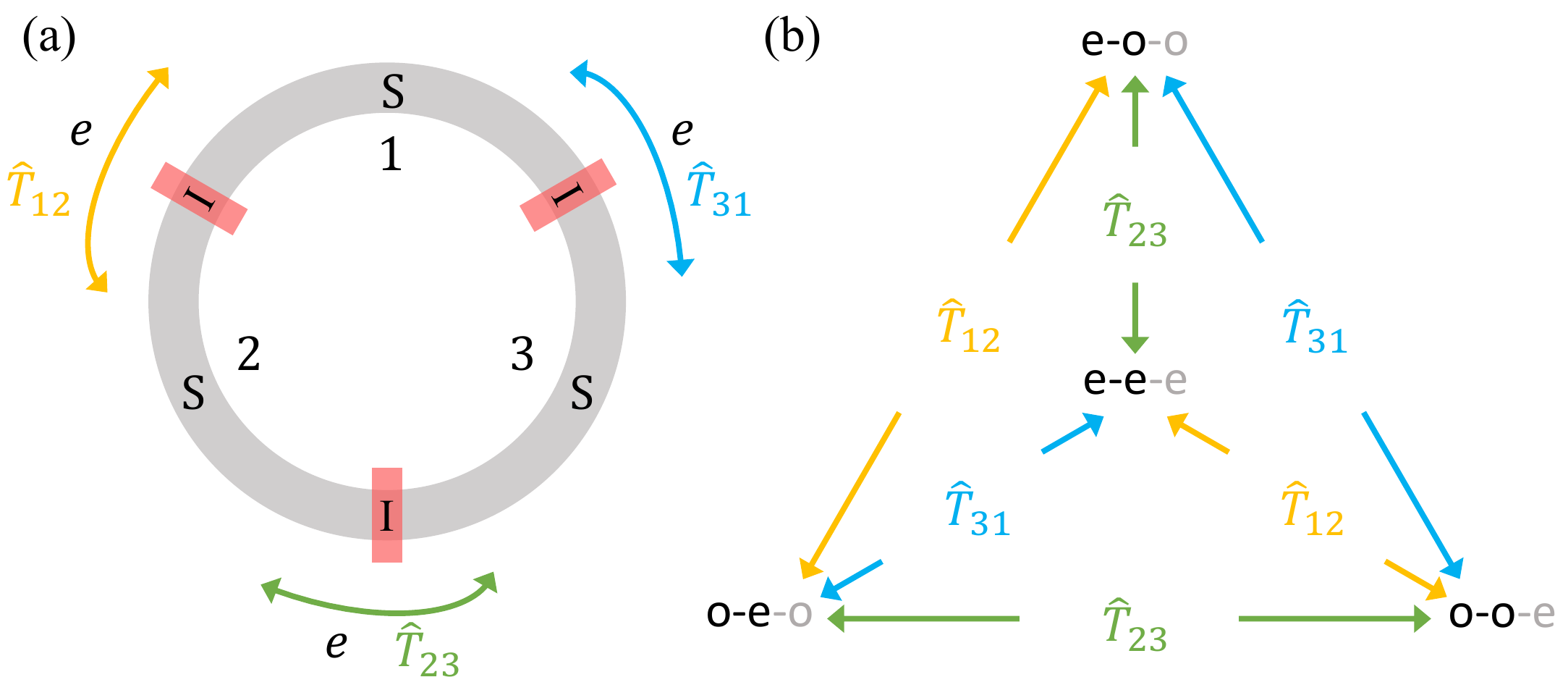}
    \caption{(a) Quasiparticle tunneling in the circulator ring. The device comprises three superconducting-insulator-superconducting (SIS) junctions across which unpaired quasiparticles can tunnel. (b) Charge-parity sectors  $ \{ \textsf{e-e(-e)}, \textsf{e-o(-o)}, \textsf{o-e(-o)}, \textsf{o-o(-e)} \}$ assuming the total charge-parity is even and tunneling operators $\hat { T}_{12} = \sin ((\hat \phi'_1 + \hat \phi'_2)/2) $, $\hat {T}_{23} = \sin(\hat \phi'_2/2) $, and $\hat{ T}_{31} = \sin (\hat \phi'_1/2) $ connecting them (see \cref{append:MEquasiparticle} for derivation of these operators).   }
    \label{fig:qpTunneling}
\end{figure}

As illustrated in \cref{fig:qpTunneling}a the circulator ring is in effect a loop of three superconducting-insulator-superconducting (SIS) junctions. Quasiparticles can tunnel across these junctions, giving rise to switching of  parities of the electron  numbers  in the superconducting islands \cite{Serniak18,Riste13}. Since the ring is capacitively isolated from outside environments, the total number of electrons is conserved.  The charge-parity configuration of the circulator ring can thus be represented by the parities of two out of the three islands, say, islands 1 and 2 only, which due to charge conservation determine the parity of the third island.
Furthermore, in the following we assume the total charge-parity of the three islands is even. Similar arguments hold for the case of an odd total charge-parity which is considered in \cref{append:OddParity}. 

The above arguments yield four accessible charge-parity sectors which we label as \mbox{\textsf{e-e-e}}, \mbox{\textsf{e-o-o}}, \mbox{\textsf{o-e-o}}, and \mbox{\textsf{o-o-e}}, where \textsf{e} denotes  even charge parity on the relevant island, and \textsf{o} denotes  odd charge parity.  Because of total charge parity conservation, the label for the third island is redundant, so for brevity we drop this label.  For example, \mbox{\textsf{e-o-o}} and \mbox{\textsf{e-o}} refer to the same charge parity sector,  
which includes  all of the charge states  satisfying $n'_1\,\mathrm{mod}\, 2 =0$ and $n'_2\, \mathrm{mod}\, 2 =1$ with $n'_1$ and $n'_2$ respectively the eigenvalues of the charge operators $\hat n'_1$ and $\hat n'_2$\footnote{Note from Eq.\ \eqref{eq:CoorTransform} that $\hat n'_1$ and $\hat n'_2$ represent charge-parities of islands 1 and 2.}. Similar definitions hold for \textsf{e-e}, \textsf{o-e}, and \textsf{o-o}.
 
 The sectors are coupled to each other by tunneling of a quasiparticle between the adjacent islands. For example,
 coupling between the sectors \textsf{e-e} and \textsf{e-o} is
 via  tunneling of a quasiparticle between islands 2 and 3. This is represented by the operator \mbox{$\hat T_{23} = \sin( (\hat \phi_3 - \hat \phi_2)/2) \equiv \sin (\hat \phi'_2/2)  $} \cite{Catelani11,Catelani12} (see \cref{append:MEquasiparticle} for derivation). 
 In \cref{fig:qpTunneling}b we illustrate all the quasiparticle-tunneling operators coupling among the four sectors. 
 
Tunneling of a quasiparticle into/out of a superconducting island is  equivalent to shifting the charge bias on that island by $\pm 1e$ \cite{Lutchyn05,Lutchyn06,Court08}, i.e.\ by half a unit charge. 
For example, if the ring is initially in the \textsf{e-e} charge sector, with charge biases $(n_{x_1},n_{x_2},n_{x_3})$, then  tunneling of a quasiparticle from island 2 to 3, will leave the ring in the sector \textsf{e-o} with  effective  charge biases   \mbox{$(n_{x_1},n_{x_2}-\tfrac{1}{2},n_{x_3}+\tfrac{1}{2})$}.   That is quasi-particle tunneling changes the charge state and therefore the effective charge bias of the islands.

This property allows us to express the Hamiltonian for all charge sectors in a self-consistent form. To do so, we order the charge basis to group  states within each of the charge-parity sectors, and in this ordered basis, the Hamiltonian matrix is block-diagonal. Each sub-block of the Hamiltonian matrix is then  given by a common functional form, $\doubleunderline{ H}^{\mathrm{ref}}(n_{x_1},n_{x_2},n_{x_3}) $, where the double-underline denotes a matrix representation of an operator expressed in the charge basis:
 \begin{align}
     \doubleunderline {H}^{\textsf{e-e}}_{\mathrm{ring}}&
     =\doubleunderline{H}^{\mathrm{ref}}(n_{x_1},n_{x_2},n_{x_3}),\nonumber\\
     \doubleunderline {H}^{\textsf{e-o}}_{\mathrm{ring}}&
     =\doubleunderline{H}^{\mathrm{ref}}(n_{x_1}, n_{x_2} + \tfrac{1}{2}, n_{x_3} - \tfrac{1}{2}),\nonumber\\
     \doubleunderline {H}^{\textsf{o-e}}_{\mathrm{ring}}&
     =\doubleunderline{H}^{\mathrm{ref}}(n_{x_1} + \tfrac{1}{2}, n_{x_2} , n_{x_3} - \tfrac{1}{2}),\nonumber\\
     \doubleunderline {H}^{\textsf{o-o}}_{\mathrm{ring}}&
     =\doubleunderline{H}^{\mathrm{ref}}(n_{x_1} + \tfrac{1}{2}, n_{x_2} - \tfrac{1}{2}, n_{x_3})\nonumber.
 \end{align}

To account for the presence of quasiparticles,
 we treat $\hat n'_1$ and $\hat n'_2$ of the Hamiltonian $\hat H_{\mathrm{ring}}$ in \mbox{Eq.\ \eqref{eq:RingHamiltonian}} as single-electron-number operators, instead of Cooper-pair-number, and the operators $\cos (\hat \phi'_1)$, $\cos (\hat \phi'_2)$, and \mbox{$\cos (\hat \phi'_1 + \hat \phi'_2)$} now describe  tunneling of two-electron charges \cite{SerniakThesis19}. In the single-electron basis $\{  \ket{n'_1,n'_2;n_0}; n'_1,n'_2 \in \mathbb{Z}  \}$, ordered to group states within each charge sector, the ring Hamiltonian $\hat H'_{\mathrm{ring}}$ is expressed as a diagonal block matrix $H'_{\mathrm{ring}}$ with four blocks corresponding to the Hamiltonians of the four sectors 
\begin{equation}
 H'_{\mathrm{ring}} =
    \begin{blockarray}{ c c c c@{\hspace{10pt}} l}
       \textsf{e-e} & \textsf{e\text{-}o} & \textsf{o\text{-}e} & \textsf{o\text{-}o}  & \\
        \begin{block}{[ c c c c@{\hspace{10pt}}]l}
              \doubleunderline H_{\mathrm{ring}}^{\textsf{e\text{-}e}} &    &    &    & \textsf{e\text{-}e} \\
              ´ & \doubleunderline H_{\mathrm{ring}}^{\textsf{e\text{-}o}} &    &    & \textsf{e\text{-}o} \\
                &    & \doubleunderline H_{\mathrm{ring}}^{\textsf{o\text{-}e}} &    & \textsf{o\text{-}e} \\
                &    &    & \doubleunderline H_{\mathrm{ring}}^{\textsf{o\text{-}o}} & \textsf{o\text{-}o}  \\
        \end{block}
    \end{blockarray}, \label{eq:RingHamiltonianBlock}
\end{equation}
where
$ \doubleunderline H_{\mathrm{ring}}^{\textsf{e\text{-}e}}$ is a matrix representation of $ \hat H_{\mathrm{ring}}^{\textsf{e\text{-}e}}$ with $n'_1$ and $n'_2$ both being even-valued and analogously for  the other  elements. Blank entries in \cref{eq:RingHamiltonianBlock} are taken to be zero.

The block structure of $\hat H'_{\mathrm{ring}}$ stems from the fact the operators $\hat n'_1$, $\hat n'_2$, $\cos(\hat \phi'_1)$, $\cos (\hat \phi'_2)$, and $\cos (\hat \phi'_1+ \hat \phi'_2)$ respect charge-parities of the ring islands, so that the ring Hamiltonian does not couple  the quasiparticle sectors.

 The quasiparticle tunneling operator \mbox{$ \hat { T}_{23} = \sin (\hat \phi'_2/2) $} couples between sector pairs (\textsf{e-e}, \textsf{e-o}) and (\textsf{o-e}, \textsf{o-o}) (as depicted in \cref{fig:qpTunneling}b), so in sector blocks it  takes the form
 \begin{equation}
T_{23} =   \begin{blockarray}{ c c c c@{\hspace{10pt}} l}
       \textsf{e-e} & \textsf{e\text{-}o} & \textsf{o\text{-}e} & \textsf{o\text{-}o}  & \\
        \begin{block}{[ c c c c@{\hspace{10pt}}]l}
            & \square &  &      & \textsf{e\text{-}e} \\
             \square  &  &  &     & \textsf{e\text{-}o} \\
              &  &  & \square     & \textsf{o\text{-}e} \\
                 &  &  \square  &  & \textsf{o\text{-}o}  \\
        \end{block}
    \end{blockarray},
 \end{equation}
where  $\square$ indicates a non-zero block sub-matrix. Similar block forms for other tunneling operators \mbox{$ \hat { T}_{12} = \sin ( (\hat \phi'_1 + \hat \phi'_2)/2) $} and \mbox{$ \hat { T}_{31} = \sin (\hat \phi'_1/2)$} are
\begin{eqnarray}
T_{12} &=&    \begin{blockarray}{ c c c c@{\hspace{10pt}} l}
       \textsf{e-e} & \textsf{e\text{-}o} & \textsf{o\text{-}e} & \textsf{o\text{-}o}  & \\
        \begin{block}{[ c c c c@{\hspace{10pt}}]l}
            &  &  & \square & \textsf{e\text{-}e} \\
 &  & \square & & \textsf{e\text{-}o} \\
  & \square &  &  & \textsf{o\text{-}e} \\
\square  &  &    &  & \textsf{o\text{-}o} \\
        \end{block}
    \end{blockarray}, \\
T_{31} &=&    \begin{blockarray}{ c c c c@{\hspace{10pt}} l}
       \textsf{e-e} & \textsf{e\text{-}o} & \textsf{o\text{-}e} & \textsf{o\text{-}o}  & \\
        \begin{block}{[ c c c c@{\hspace{10pt}}]l}
             &  & \square &  & \textsf{e\text{-}e} \\
 &  &  & \square & \textsf{e\text{-}o} \\
 \square &  &  &  & \textsf{o\text{-}e} \\
  & \square &    &  & \textsf{o\text{-}o}\\ 
        \end{block}
    \end{blockarray}.
\end{eqnarray}

\subsection{Fluctuations between charge-parity sectors} \label{subsec:QPFluctuations}

Having identified the four charge-parity sectors, here we evaluate the transition rates between them and compute their respective  circulation. To this end, we derive the master equation for the ring density operator $\rho'$ in the presence of quasiparticle tunneling (see \cref{append:MEquasiparticle} for derivation)
\begin{eqnarray}
\dot \rho' (t) &=& -i [\hat H'_{\mathrm{ring}} -i \sqrt{\Gamma} \sum_{j} (\beta_j e^{-i \omega_{d} t } \hat q_{j,+} - \mathrm{H.c.} ), \rho'(t) ] \nonumber \\
    && + \sum_{j}\sum_{s}\sum_{k>k'} \Gamma^{(j)}_{k,s;k',s} \mathcal{D}[\ket{k',s}\!\bra{k,s}] \rho' (t) \nonumber \\
&& + \sum_{j\ne j'} \sum_{s,s'}\sum_{k,k'} \Gamma^{(jj')}_{k,s;k',s'} \mathcal{D}[\ket{k',s'}\!\bra{k,s}] \rho'(t), \qquad \quad \label{eq:MEqp}
\end{eqnarray}
where $j$ and $j'$ index the islands,
$s$ and $s'$ label the quasiparticle sectors $\{ \textsf{e\text{-}e},\textsf{e\text{-}o},\textsf{o\text{-}e},\textsf{o\text{-}o}\}$, $k$ and $k'$ index the ring eigenstates, and $\ket{k,s}$ denotes a ring eigenstate $\ket{k}$ in the sector $s$. In Eq.\ \eqref{eq:MEqp} the second line describes the inner-sector relaxation transition due to couplings to the waveguides with the rate \mbox{ $\Gamma^{(j)}_{k,s;k',s}=  \Gamma |\langle k',s| \hat q_j |k,s \rangle|^2$}, whereas the third line
describes  the inter-sector jump (i.e., quasiparticle tunneling) from a state $\ket{k,s}$ in the sector $s$ to another state $\ket{k',s'}$ in the sector $s'$ 
with the rate $\Gamma^{(jj')}_{k,s; k',s'}$ given by \cite{CatelaniPRB11}
\begin{equation}
\Gamma^{(jj')}_{k,s; k',s'} = \big| \langle k',s' | \hat T_{jj'} | k,s \rangle \big|^2 S_{\mathrm{qp}} (\omega_{k,s;k',s'}),
\end{equation}
where the sector-coupling operator $ \hat{ T}_{jj'}$ is  given explicitly in \cref{fig:qpTunneling}b for each inter-sector transition, $\omega_{k,s;k',s'} $ is the transition energy between the states $\ket{k,s}$ and $\ket{k',s'}$, and $S_{\mathrm{qp}} (\omega) $ is the quasiparticle spectral density. For  a relaxation process with $\omega>0$, $S_{\mathrm{qp}} (\omega) $  is given by \cite{Catelani11,CatelaniPRB11}
\begin{eqnarray}
S_{\mathrm{qp}} (\omega) &=& \frac{16 E_J}{\pi} \int_{0}^{\infty} dx \frac{1}{\sqrt{x} \sqrt{x+\omega/\Delta}} \big( f[(1+x)\Delta] \nonumber \\
&& \times \{1-f[(1+x)\Delta + \omega] \} \big), \label{eq:qpSpectralDensity}
\end{eqnarray}
where $f[E] $ is the quasiparticle distribution function. At equilibrium, one would expect that $f[E]$ is of the form $  1/(\exp(E/k_BT)+1) $  \cite{Catelani14}, but non-equilibrium quasiparticles may be present modifying $f[E]$ \cite{SerniakThesis19}. For an excitation process with $\omega<0$, in Eq.\ \eqref{eq:qpSpectralDensity} we make replacements $x \to x- \omega/\Delta$ and $\omega \to - \omega$.

At equilibrium and in the limit of high frequency, $\delta E \ll \omega \ll \Delta$ with $\delta E$ the characteristic energy of quasiparticles \cite{Catelani11,CatelaniPRB11}, we can approximate \mbox{$S_{\mathrm{qp}}(\omega) = (8E_J/\pi) \sqrt{2\Delta/\omega}\, x_{\mathrm{qp}} $}, where $ x_{\mathrm{qp}}$ is the quasiparticle density normalized by the Cooper-pair density 
\begin{equation}
    x_{\mathrm{qp}} = \sqrt{2\pi k_B T/\Delta} e^{-\Delta/k_B T}. \label{eq:qpDensity}
\end{equation}
At $ T \approx 20 \, \mathrm{mK}$ and for aluminum superconductors with $\Delta \approx 1.76\, k_B T_c $ and $T_c \approx 1.35\, \mathrm{K}$ \cite{SerniakThesis19}, $x_{\mathrm{qp}}$ should be of order $10^{-53}$, effectively suppressing quasiparticle tunneling in equilibrium BCS superconductors.
Experimentally observed results for superconducting circuits nonetheless showed that $x_{\mathrm{qp}} \approx 10^{-8} - 10^{-6} $ \cite{Martinis09,Vool14,Wang14}. This indicates a small but non-negligible population of non-equilibrium quasiparticles. The origin of these is not certain \cite{Mannila21}, but may arise from stray photons \cite{Serniak18,Houzet19}, ionizing radiation from surrounding radioactive materials \cite{Vepsalainen20,Cardani20}, and cosmic rays \cite{Martinis09}. 
Further,  electrons and  photon baths can be out of equilibrium, so that  electrons are typically hotter than the base fridge temperature \cite{Borzenets13}. In any case, we implicitly assume an empirical value for $x_{\mathrm{qp}}$.

 To include non-equilibrium quasiparticles, in \cref{eq:qpDensity} we replace the base temperature $T$ ($\approx 20\, \mathrm{mK}$) with an effective quasiparticle one $T_{\mathrm{qp}} \approx 200\, \mathrm{mK}$ \cite{Serniak18,SerniakThesis19}. For $E_J \sim 2\pi \times 10\, \mathrm{GHz}$ and $E_{C_\Sigma}/E_J = 0.35 $  we numerically find \mbox{$\big| \langle k',s' | \hat{ T}_{jj'} | k,s \rangle \big|^2 \sim 10^{-2} - 10^{-1}$} and \mbox{$\omega_{k,s; k',s'} \sim 2\pi \times 10 \, \mathrm{GHz}$}, so that for $\Delta \approx 1.76\, k_B T_c$ with $T_c \approx 1.35 \, \mathrm{K}$ the quasiparticle induced transition rate $\Gamma^{(jj')}_{k,s; k',s'}$ will be of order  $ 0.1 - 1 \, \mathrm{kHz}$. This corresponds to quasiparticle lifetime of order $0.1 - 1\, \mathrm{ms}$ as observed in several experiments \cite{Riste13,Ferguson06,Court08,Lenander11}. The quasiparticle temperature $T_{\mathrm{qp}}$ determines the quasiparticle-tunneling rate (i..e, the sector fluctuation rate) and does not affect the circulation performance of each quasiparticle sector. In the absence of quasiparticle fluctuations, the parameters would be chosen to optimize the fidelity in the fixed quasiparticle sector. Quasiparticle-trapping techniques such as normal metal traps \cite{Joyez94,Court08,Martinis21} and gap engineering \cite{Aumentado04,Sun12,Kalashnikov20} can be used to suppress non-equilibrium quasiparticle population. This effectively reduces the quasiparticle temperature $T_{\mathrm{qp}}$ and subsequently the quasiparticle-tunneling rate. A long period free of quasiparticle tunneling events in the circulator will benefit device calibration and make the circulation performance stable.

\subsubsection{Symmetric Josephson-junction ring}

\begin{figure}[b!]
\centering
 \includegraphics[scale=0.87]{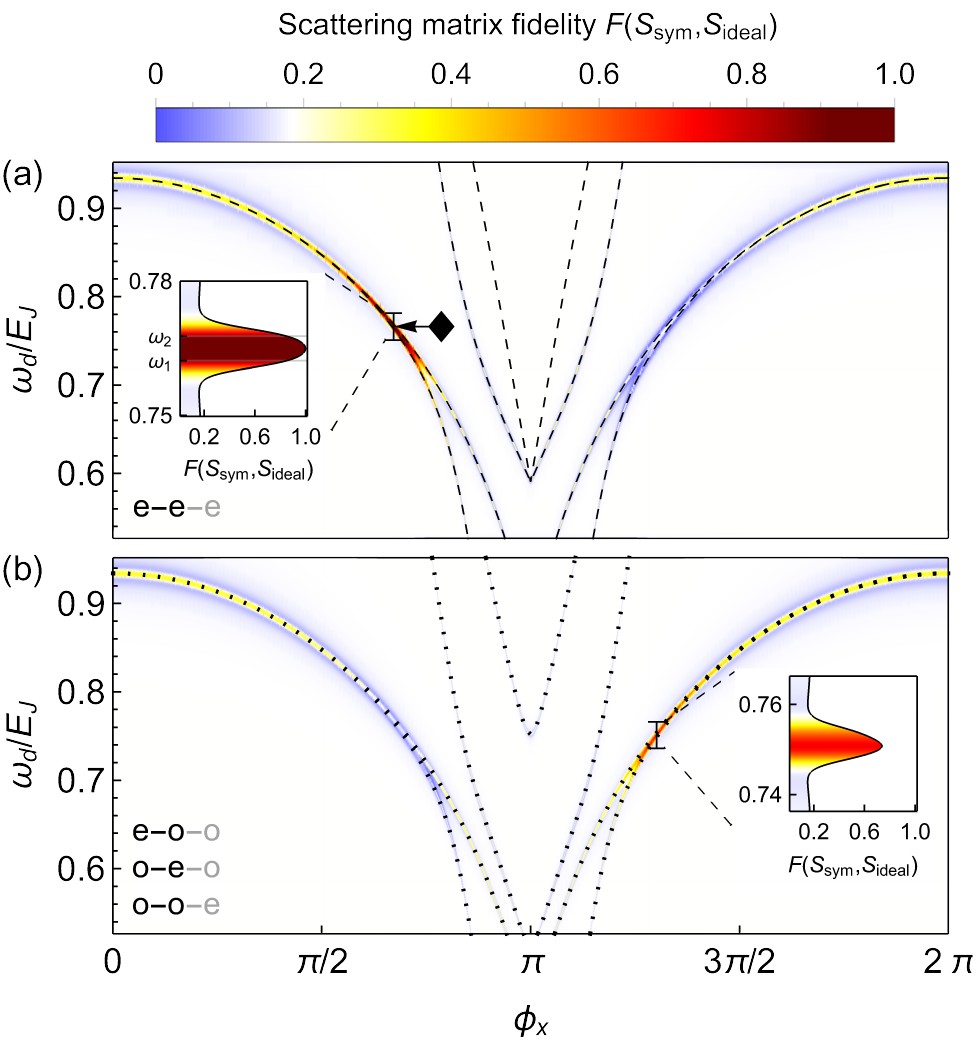}
\caption{ Dependence on the nondimensional flux bias $\phi_x$ and the driving frequency $\omega_d$ of the scattering-matrix fidelity $F(S_{\mathrm{sym}},S_{\mathrm{ideal}})$ for a symmetric circulator ring within (a) the sector \textsf{e-e(-e)} and (b) the sectors \textsf{e-o(-o)}, \textsf{o-e(-o)}, and \textsf{o-o(-e)}. Here owing to the junction and charge-bias symmetries the scattering-matrix fidelities within the three sectors \textsf{e-o}, \textsf{o-e}, and \textsf{o-o} are exactly the same. The insets in both panels show the bandwidth of the circulator at a fixed value of $\phi_x$ that yields a highest fidelity. The dashed lines in panel (a) indicate the transition energies between the ground state of the ring and different excited-states in the quasiparticle sector \textsf{e-e}. The $\blacklozenge$ symbol indicates the optimal working point. The dotted lines in panel (b) indicate transition energies in the sector \textsf{e-o} (which is the same as in \textsf{o-e} and \textsf{o-o} for a symmetric device). Relevant parameters are the same as in \cref{fig:OptFideSym} but with $n_{x_j}=1/3$ ($j=1,2,3$). We also note that for illustration purposes here and in \cref{fig:ASymSectorFluctuation} we use a higher waveguide impedance $Z_{\mathrm{wg}} = 200\, \Omega$ to increase the coupling strength $\Gamma$. }
\label{fig:SymSectorFluctuation}
\end{figure}

Since its Hamiltonian is block-diagonal across the charge-parity sectors,  the circulator ring will evolve within one particular sector, with intermittent, incoherent transitions between the sectors when quasiparticle tunneling events occur. Accordingly, in \cref{append:MEquasiparticle} we unravel the master equation in Eq.\ \eqref{eq:MEqp} into a stochastic jump evolution equation \cite{Stace03,GardinerBook04} with intermittent inter-sector jumps. This allows us to compute the circulation performance of the device within a given quasiparticle sector during intervals in which no quasiparticle jumps occur. We note that the quasiparticle tunneling rate $\Gamma^{(jj')}_{k,s;k',s'}\, (\sim 1\, \text{kHz})$ is much smaller than the waveguide coupling $\Gamma\, (\sim 100\, \text{MHz})$, so $\Gamma^{(jj')}_{k,s;k',s'}$  has a negligible effect on the spectra within each quasiparticle sector. The principle effect of the quasiparticle tunneling terms is just to drive transitions between sectors, with an inter-sector transition rate given by $\Gamma^{(jj')}_{k,s;k',s'}$. 
In what follows we compare the circulation performance in each quasiparticle sector, considering both ideal symmetric rings, and realistic asymmetric  rings in which the junctions are not identical.

In \cref{fig:SymSectorFluctuation} we show variation of the scattering-matrix fidelity $F(S_\mathrm{sym}, S_{\mathrm{ideal}} )$ as a function of the reduced external flux $\phi_x$ and the driving frequency $\omega_d$ for a symmetric circulator ring  with symmetric charge biases ($n_{x_j}=1/3$ for $j=1,2,3$) in the four quasiparticle sectors. We note that here for illustration purposes we increase the coupling strength $\Gamma$ defined in \cref{eq:CouplingExpression} by effectively choosing a higher waveguide impedance $Z_{\mathrm{wg}} = 200\, \Omega$; smaller $Z_{\mathrm{wg}}$ reduces the bandwidth proportionally.

For the sector \textsf{e-e} in \cref{fig:SymSectorFluctuation}a we observe that a high-fidelity region (dark-red) and a low-fidelity region (blue) are symmetric about $\phi_x = \pi$. Such symmetry is owing to the mirror-symmetry of the eigenstates of the circulator ring with respect to a half-quantum flux bias, by which the high-fidelity region yields strong clockwise signal circulation while the low-fidelity region yields counter-clockwise signal circulation (see also Ref.\ \cite{Muller18}). 

The optimal working point in  \cref{fig:SymSectorFluctuation}a is found at \mbox{$(\phi_{x }^\blacklozenge,\omega_{d}^\blacklozenge)\approx(2.11,0.77 E_J)$} (labeled by the  $\blacklozenge$ symbol), in which the driving frequency $\omega_{d}^\blacklozenge$ lies in between the first two excited-state eigenenergies $\omega_1$ and $\omega_2$, as expected from the condition in Eq.\ \eqref{eq:driveCondition}. Around this optimal working point,  the bandwidth  evaluated from the inset in  \cref{fig:SymSectorFluctuation}a is around $0.01 E_J$. For $E_J \sim 2\pi \times 10\, \mathrm{GHz}$, the bandwidth is $2\pi \times 100 \,\mathrm{MHz}$. This is consistent with the estimate made from \cref{eq:CouplingExpression}, which yields the waveguide coupling strength $\Gamma \sim 0.01 \, E_J $ for $C_c/C_{\Sigma} \approx 0.31 $, $Z_{\mathrm{wg}} = 200\, \Omega$, $\omega_d \approx 0.77 E_J$ (see \cref{append:SimuPara} for detailed parameter values used in simulations).

We note that a background circulation fidelity at about $0.18$ for a non-circulating device is indicated by white in the color scale in \cref{fig:SymSectorFluctuation} (and \cref{fig:ASymSectorFluctuation} as well).
This background fidelity is present due to the fact that when the driving frequency is far off-resonant with respect to the excited-state energies,
transmission of signals in the circulator is very small and there is only reflection.  In particular, in  \cref{eq:Selement} when \mbox{$\Delta \omega_k/\Gamma \gg \gamma_k, |\langle k | \hat q_j | 0\rangle \langle 0 | \hat q_i | k\rangle | $} due to large detuning $\Delta \omega_k$, the second part of \cref{eq:Selement} is close to zero, rendering the scattering matrix very similar to an identity matrix. In this case one finds $F(|S|, S_{\mathrm{ideal}} ) \simeq F(\mathbb{1}, S_{\mathrm{ideal}} ) \approx 0.18 $.

In the other quasiparticle sectors \textsf{e-o}, \textsf{o-e}, and \textsf{o-o},  the scattering-matrix fidelities are exactly identical for a symmetric circuit and are shown in \cref{fig:SymSectorFluctuation}b. 
Comparing  Figs. \ref{fig:SymSectorFluctuation}a and \ref{fig:SymSectorFluctuation}b, the locations of the high-fidelity and low-fidelity regions are exchanged. 
Concretely,
at \mbox{$(\phi_{x }^\blacklozenge,\omega_{d}^\blacklozenge)$} in \cref{fig:SymSectorFluctuation}a,  we find \mbox{$F(S_\mathrm{sym}, S_{\mathrm{ideal}} ) \approx 0.99$} and \mbox{$(S_{21},S_{31}) \approx (0.001, 0.996)$}. In contrast, at the same working point in \cref{fig:SymSectorFluctuation}b we find
\mbox{$F(S_\mathrm{sym}, S_{\mathrm{ideal}} ) \approx 0.14$} and \mbox{$(S_{21},S_{31}) \approx (0.622,0.274)$}. These indicate significant reverse of signal circulation from clockwise to counter-clockwise 
and the adverse influence of quasiparticle tunneling. Assuming the device is circulating signals clockwise at the high-fidelity region in the sector \textsf{e-e} as in \cref{fig:SymSectorFluctuation}a, then an event of tunneling of a quasiparticle suddenly transforms the circulator to the other sectors and reverses the circulation direction as in \cref{fig:SymSectorFluctuation}b.

\subsubsection{Asymmetric Josephson-junction ring}

\begin{figure}[h!]
\centering
 \includegraphics[scale=0.87]{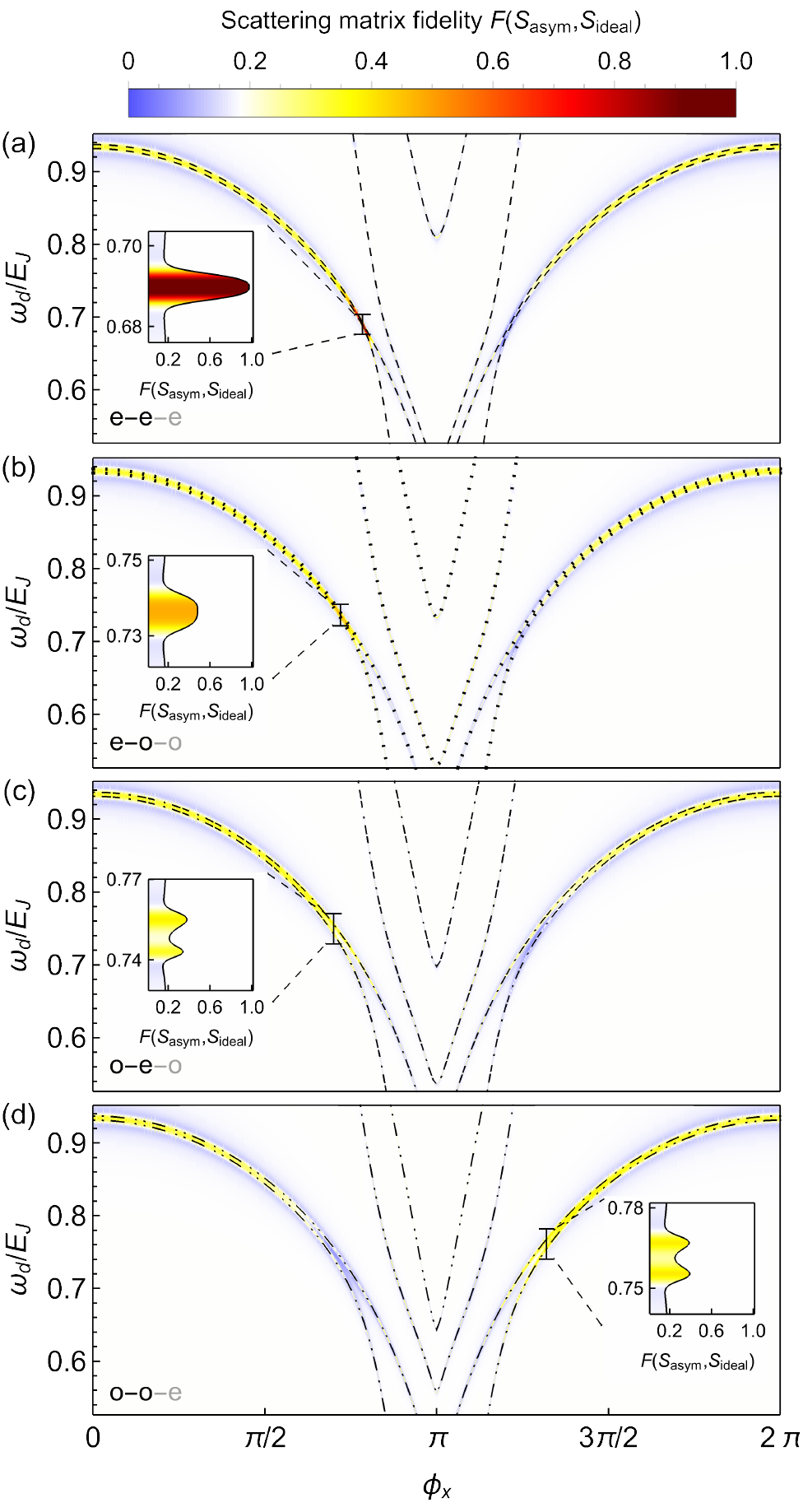}
\caption{Dependence on the nondimensional flux bias $\phi_x$ and the driving frequency $\omega_d$ of the fidelity of the scattering matrix $F(S_{\mathrm{asym}},S_{\mathrm{ideal}})$ for an asymmetric circulator ring within (a) the sector \textsf{e-e(-e)} and (b) the sector \textsf{e-o(-o)}, (c) the sector \textsf{o-e(-o)}, and (d) the sector \textsf{o-o(-e)}. The insets in all panels show the bandwidth of the circulator at a fixed value of $\phi_x$ that yields a highest fidelity. Ground-to-excited-state transition energies of the circulator ring in each quasiparticle sector are also plotted (dashed, dotted, dot-dashed, and dot-dot-dashed for the four sectors, respectively). Relevant parameters are the same as in \cref{fig:OptFideASym} but for illustration purposes we choose $Z_{\mathrm{wg}} = 200\, \Omega$ to increase the coupling strength, which results in $\delta E_J/\Gamma \sim 1$ and subsequently an optimized fidelity in panel (a) near 1 consistent with the analysis in \cref{fig:checkASYM}b.}
\label{fig:ASymSectorFluctuation}
\end{figure}

We consider the same junction asymmetry as in \cref{subsec:AsymNum} with $E_{J_1} /E_{J}=1$, $E_{J_2}/E_J = 1.01$, and $E_{J_3}/E_J=0.99$. We numerically optimize the scattering-matrix fidelity $F(S_{\mathrm{asym}}, S_{\mathrm{ideal}} )$ for the sector \textsf{e-e} and find its optimal value ($\approx 0.97$) at $(\phi_x, \omega_d, n_{x_1},n_{x_2},n_{x3}) = (2.46, 0.69\, E_J, 0.10, 0.19, 0.84)$. The optimal working points for the other sectors are obtained by shifting the relevant charge biases by half of a Cooper pair while keeping the external flux and the driving frequency the same. For example, for the sector \textsf{e-o} we find the optimal fidelity (also $\approx 0.97$) at $(\phi_x, \omega_d, n_{x_1},n_{x_2},n_{x3}) = (2.46, 0.69 E_J, 0.10, 0.19-1/2, 0.84+1/2)$, and similarly for the sectors \textsf{o-e} and \textsf{o-o}. In what follows, we fix the charge bias configuration at $(n_{x_1},n_{x_2},n_{x3}) = (0.10, 0.19, 0.84)$ to be the same for all sectors, corresponding to the experimental reality that we assume the circulator to be in the sector \textsf{e-e} at all times; quasiparticles will therefore degrade the performance. We plot the scattering fidelity versus $\phi_x$ and $\omega_d$ for the four sectors in \cref{fig:ASymSectorFluctuation}.

Figure \ref{fig:ASymSectorFluctuation} shows that the four sectors have quite different performances. 
The sectors \textsf{e-e}, \textsf{e-o}, and \textsf{o-e} share the same high-fidelity region with $2.1 \le \phi_x \le 2.5$ but with decreasing efficiencies (see the insets in Figs.\ \ref{fig:ASymSectorFluctuation}a-c), while the sector \textsf{o-o} has its high-fidelity region mirror-flipped compared to those in the other sectors. 
This is different from the case of a symmetric circuit considered previously which exhibits exchange of the high-fidelity and low-fidelity regions in the sector \textsf{e-e} and the other  sectors.
Such a difference is a result of junction asymmetry $\delta E_J = 0.01 E_J$ making the sectors \textsf{e-o}, \textsf{o-e}, and \textsf{o-o} no longer equivalent as in the symmetric-circuit case. 
Quasiparticle-tunneling induced jumps between these different sectors will make the circulator operate unreliably.

\subsection{Composition of quasiparticle spectra}

\begin{figure}[ht!]
\centering
 \includegraphics[scale=0.661]{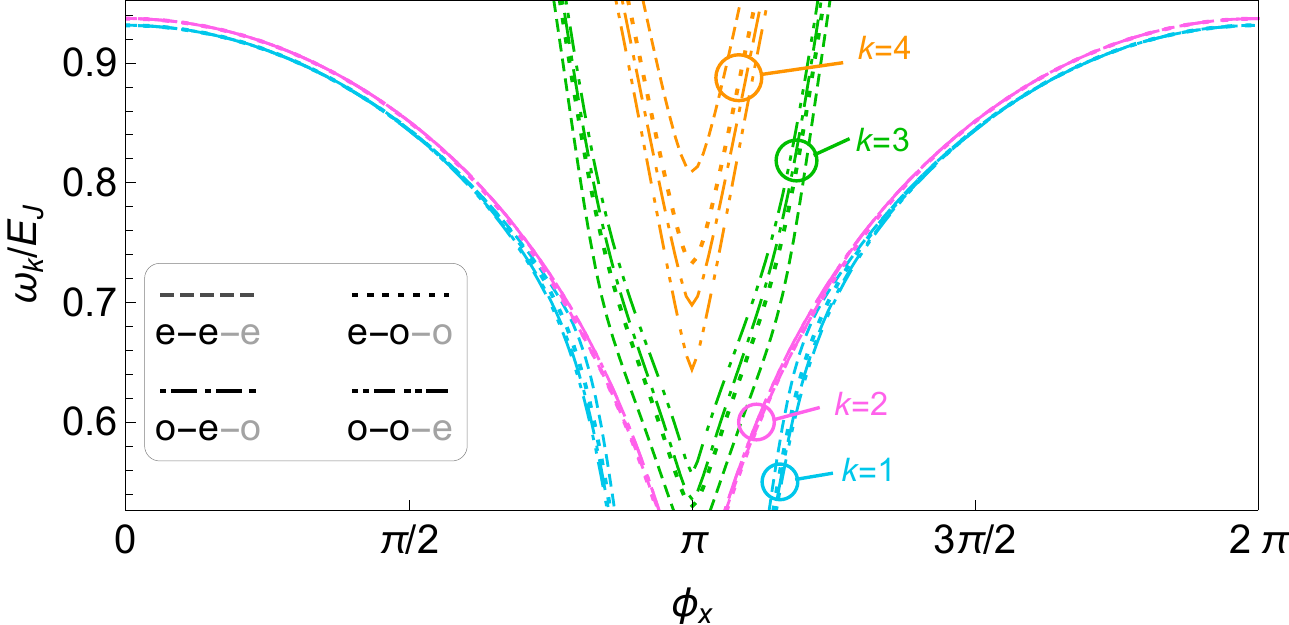}
\caption{Composition of the ground-to-excited-state transition energies for the first four excited states $\omega_k$ ($k=1,2,3,4$ from bottom to top with different colors) from each of the four quasiparticle sectors. The line styles of the sector spectra here match those in \cref{fig:ASymSectorFluctuation}. Relevant parameters are the same as in \cref{fig:ASymSectorFluctuation}.  }
\label{fig:sectorOverlap}
\end{figure}

Since circulation in our system is a resonant effect, reflection or transmission measurements of the circulator ring will reveal its energy spectrum. Each quasiparticle sector has a distinct spectrum, so   measurements performed on a timescale longer than quasiparticle lifetimes will show all the spectra from  the four  sectors superimposed.
Coexistence of the even and odd sectors has been observed in experiments with the single-Cooper-pair transistor and the Cooper-pair-box/transmon qubit \cite{Aumentado04,Serniak18,Sun12,Riste13} that feature the ``eye-pattern"  composing of both even and odd transitions. 

In \cref{fig:sectorOverlap} we show the four sector spectra with the
first four excited-state energies $\omega_k$ ($k=1,2,3,4$ from bottom to top with different colors) as functions of the reduced external flux $\phi_x$ for an asymmetric circulator ring.  We numerically compute the eigenenergies of the Hamiltonian, \cref{eq:RingHamiltonianBlock} for each quasiparticle sector (i.e.,  we diagonalize the closed-system as for the dashed lines in \cref{fig:ASymSectorFluctuation} disregarding waveguide or quasiparticle induced decoherence, and then compose the eigenenergies together in a single plot).  
The multi-sector spectra in \cref{fig:sectorOverlap} serve  as a signature of the presence of the different quasiparticle sectors when carrying out initial spectroscopic measurements on the circulator ring. They may also be a map to distinguish the different quasiparticle sectors.

\section{Discussion}

Based on the results in the previous sections, we can sketch a three-step procedure to calibrate the Josephson-junction-ring circulator as follows.

(i) First,  we perform transmission or reflection measurement of the ring device to obtain its spectrum, from which we fit its energy scales (i.e., $E_{C_\Sigma}$ and $E_{J_j}$). We also look for any spectra overlap due to quasiparticle tunneling as in \cref{fig:sectorOverlap}. No signs of such overlap indicate a device free of quasiparticles, which will facilitate finding the optimal working points.  

(ii) Second, given the ring energy parameters we numerically estimate the three charge biases, the external flux, and the driving frequency to achieve optimal circulation as done  in \cref{sec:AEPredictions} and \cref{sec:Optimization}.

(iii) Third, we incorporate a particular optimization routine into measurements of the device and optimize the scattering matrix fidelity around the parameters found in step (ii).

Of particular importance is  to take into account junction asymmetry. As analyzed in \cref{sec:Optimization} the tolerance level of signal circulation to junction asymmetry is set by the coupling strength and in general the smaller the ratio between junction asymmetry and the coupling strength the better the scattering matrix fidelity. SQUID-geometry junctions with tunable Josephson energies can be used to mitigate asymmetry imperfection, but might complicate device operation and introduce new noise channels. Alternatively, one can increase the coupling strength by employing high-impedance waveguides \cite{Andersen17,Stockklauser17} to relax constraints on device fabrication as well as enhancing the working bandwidth.

We suggest a set of fabrication parameters for the ring circulator: $E_{C_\Sigma}/2\pi \sim 3\, \mathrm{GHz} $, $\bar{E}_J/2\pi \sim  10\, \mathrm{GHz}$, $ \Gamma/2\pi \sim 100\, \mathrm{MHz} $, and $|\delta E_J| \lesssim 2 \Gamma $, by which $|\delta E_J|/\Gamma  \lesssim 2$ allowing us to achieve a high circulation fidelity [see \cref{fig:checkASYM}] with an operational bandwidth around $2\pi \times 100\, \text{MHz}$. Besides,
quasiparticle-trapping techniques such as normal metal traps \cite{Joyez94,Court08,Martinis21} and gap engineering \cite{Aumentado04,Sun12} combined with careful shielding \cite{Bell14} can be harnessed to suppress unpaired quasiparticles, with the goal of creating a period free of quasiparticles
 at a timescale [about seconds or minutes \cite{Kalashnikov20,Mannila21}] much larger than the needed optimization time [at the order of milliseconds as estimated in \cref{sec:Optimization}]. Conditioned on this, we expect to apply the optimization procedure presented here to a real experimental setup to calibrate the circulator device.

\section{Conclusion} \label{sec:Conclusion}

The passive on-chip superconducting circulator proposed in Ref.\ \cite{Muller18} is intriguing, as it operates passively and may facilitate scaling up  superconducting circuit experiments. Practical operation of this device necessitates consideration of two challenging issues: tuning the various external control parameters to the optimal working points and reducing the impact of parameter instabilities. Here we have shown that even with a simple optimization routine the multi-parameter optimization can be implemented quickly within less than $50$ optimization steps to determine the desired operating points for the circulator device. The optimization is supported by our semi-analytic treatment of the scattering problem which elucidates intuition for numerical results.  

As for parameter instabilities, we have considered a detrimental type of charge noise, that is, quasiparticle tunneling. We find that  tunneling of quasiparticles across the circulator junctions creates four available operating sectors, differing in the charge parities of the superconducting islands. Under the same working conditions, each sector circulates signals differently with varied circulation direction and efficiency. Stochastic jumps between the sectors due to quasiparticle tunneling subsequently may render the circulator performance inefficient. We suggest using quasiparticle-trapping and shielding techniques to reduce quasiparticle population and the quasiparticle-tunneling rate, thus potentially rendering the device unaffected by quasiparticles for a period much longer than the required optimization time.

\begin{acknowledgments}
We thank Andr\'es Rosario Hamann for useful discussions.
This research was supported by the Australian Research Council Centres of Excellence for Engineered Quantum Systems (Projects No.\ CE170100009 and No.\ CE110001013) and the Swiss National Science Foundation through NCCR Quantum Science and Technology.  
\end{acknowledgments}

\appendix

\section{Circuit quantization} \label{append:CircuitQuantization}

\subsection{Combined circulator-waveguide system}

\begin{figure}[h]
    \centering
    \includegraphics[scale=0.7]{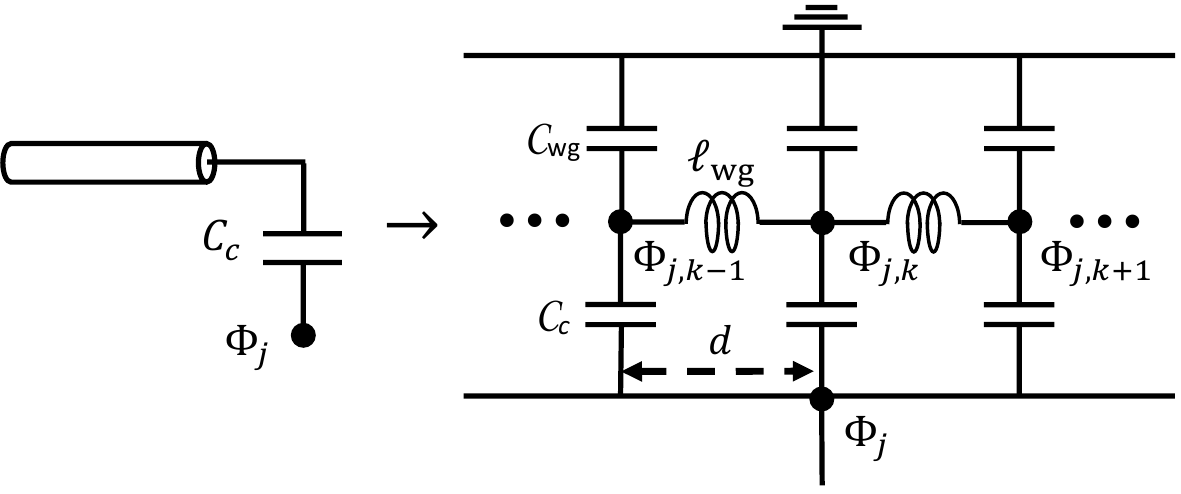}
    \caption{(left) Formal representation of the waveguide-circulator capacitive coupling and (right) microscopic model for this coupling in which the coupling capacitance is treated as a distributed element \cite{Le19}.}
    \label{fig:LumpedCircuitCoupling}
\end{figure}

Figure \ref{fig:LumpedCircuitCoupling} shows the formal representation (left) and the microscopic model (right) of the coupling between the ring island $j$ represented by the canonical flux $\Phi_j$ and the waveguide $j$ in which the coupling capacitance $C_c$ and the waveguide are treated as  distributed elements. The waveguide of length $L$ is decomposed into unit cells of length $d$, so that the total unit-cell number is $N=L/d$. The waveguide capacitance and inductance per unit cell are $\mathcal C_{\mathrm{wg}}$ and $ \ell_{\mathrm{wg}}$, yielding the waveguide capacitance and inductance per unit length as $\bar{\mathcal C}_{\mathrm{wg}} = \mathcal C_{\mathrm{wg}}/d$ and $\bar{\ell}_{\mathrm{wg}} = \ell_{\mathrm{wg}}/d$.  The coupling capacitance $C_{c}$ has  length $y$ spreading across $D = y/d$ waveguide unit cells, by which the coupling capacitance per unit cell is $\mathcal{C}_c = C_{c}/D$. We also assume that the waveguide length is much larger than that of the coupling capacitance, i.e., $ L \gg y $.

We perform circuit quantization for the combined circulator-waveguide system assuming  symmetric Josephson junctions, $E_{Jj} = E_J$ and $C_{Jj}=C_J$, symmetric gate capacitances, $C_{x_j}=C_x$, and symmetric coupling capacitances, $C_{c_j} = C_c$. The Lagrangian of the total circuit in the absence of external biases is 
\begin{eqnarray}
\mathcal L_{\mathrm{tot}} &=& \sum_{j=1}^{3} \frac{C_J}{2} (\dot \Phi_{j+1} - \dot \Phi_j)^2 + \frac{C_x}{2}  \dot \Phi_j^2  \nonumber \\
 && +  E_J \cos \bigg( \frac{2\pi}{\Phi_0} (\Phi_{j+1} - \Phi_j) \bigg) \nonumber \\
 && + \sum_{j=1}^{3} \sum_{k=1}^{N} \frac{\mathcal C_{\mathrm{wg}}} {2} \dot \Phi_{j,k}^2 - \frac{(\Phi_{j,k+1} - \Phi_{j,k})^2}{2 \ell_{\mathrm{wg}} } \nonumber \\
 && + \sum\limits_{j=1}^{3} \sum_{k=1}^{D} \frac{\mathcal C_c}{2} (\dot \Phi_{j,k} - \dot \Phi_j )^2,
\end{eqnarray}
where the first and second lines, respectively, represent the Lagrangians of the circulator ring and the waveguides in the absence of any coupling and the third line represents their capacitive coupling. We decompose $\mathcal L_{\mathrm{tot}}$ into
\begin{equation}
    \mathcal L_{\mathrm{tot}} = \mathcal L_{\mathrm{ring}} + \mathcal L_{\mathrm{wg}} + \mathcal L_{\mathrm{int}},
\end{equation}
where
\begin{eqnarray}
 \mathcal L_{\mathrm{ring}} &=& \frac{1}{2} \mathbf{\dot \Phi} \mathbb{C} \mathbf{\dot \Phi} +  E_J \sum_{j=1}^3 \cos \bigg( \frac{2\pi}{\Phi_0} (\Phi_{j+1} - \Phi_j) \bigg), \qquad \label{eq:ringLagrangian} \\
 \mathcal L_{\mathrm{wg}} &=& \sum_{j=1}^{3}  \sum_{k=1}^{N} \frac{\mathcal C_{\mathrm{wg}}} {2} \dot \Phi_{j,k}^2  - \frac{(\Phi_{j,k+1} - \Phi_{j,k})^2}{2 \ell_{\mathrm{wg}} } \nonumber \\
&&  + \sum_{j=1}^3 \sum_{k=1}^D \frac{\mathcal C_c}{2} \dot \Phi^2_{j,k}, \label{eq:waveguideLagrangian} \\
 \mathcal L_{\mathrm{int}} &=& - \sum\limits_{j=1}^{3} \sum_{k=1}^{D} \mathcal C_c \dot \Phi_{j,k}  \dot \Phi_j, \label{eq:interactionLagrangian}
\end{eqnarray}
where $\mathbf{\dot \Phi} = \{\dot \Phi_1, \dot \Phi_2, \dot \Phi_3\}$ and 
\begin{equation}
    \mathbb{C} = \left( \begin{array}{ccc}
      C_{\Sigma} - C_J & - C_J & - C_J    \\
       -C_J & C_{\Sigma} - C_J & - C_J  \\
       -C_J & -C_J & C_{\Sigma} - C_J
    \end{array} 
    \right),
\end{equation}
with 
\begin{equation}
    C_{\Sigma} =  3C_J + C_x + C_c.
\end{equation}

 We determine the conjugate momenta $\{Q_j, Q_{j,k}; j=1,2,3 ; k=1, \dots, N \}$ via the equations $Q_j = \partial \mathcal L_{\mathrm{tot}}/\partial \dot \Phi_j$ and $Q_{j,k} = \partial \mathcal L_{\mathrm{tot}}/\partial \dot \Phi_{j,k}$ and perform Legendre transformation to compute the (classical) Hamiltonian $\mathcal H_{\mathrm{tot}}$. Keeping terms to the first order of $\mathcal C_c/C_{\mathrm{others}}$ only (that is, we assume that the unit-cell coupling capacitance $\mathcal C_c$ is much smaller than other capacitances), we find
\begin{equation}
    \mathcal H_{\mathrm{tot}} = \mathcal H_{\mathrm{ring}} + \mathcal{H}_{\mathrm{wg}} + \mathcal H_{\mathrm{int}}.
\end{equation}
Concretely, $\mathcal H_{\mathrm{ring}}$ is given by
\begin{equation}
\mathcal H_{\mathrm{ring}} = \frac{1}{2} \mathbf{Q} \mathbb{C}^{-1} \mathbf{Q}  - E_J \sum_{j=1}^3 \cos \bigg( \frac{2\pi}{\Phi_0} (\Phi_{j+1} - \Phi_j) \bigg),
\end{equation}
where $\mathbf{Q} = \{Q_1, Q_2, Q_3\}$.  
The waveguide Hamiltonian $\mathcal H_{\mathrm{wg}}$ is
\begin{equation}
    \mathcal H_{\mathrm{wg}} = \sum_{j=1}^{3} \sum_{k=1}^{N} \frac{Q^2_{j,k}}{2 \mathcal C_{\mathrm{wg}}} + \frac{(\Phi_{j,k+1} - \Phi_{j,k})^2}{2\ell_{\mathrm{wg}}} .
\end{equation}
Lastly, the interaction $\mathcal H_{\mathrm{int}}$ is given by
\begin{eqnarray}
     \mathcal H_{\mathrm{int}} & = &  \frac{\mathcal C_c}{\mathcal C_{\mathrm{wg}} (C_x +C_c) C_{\Sigma}}  \sum_{j=1}^{3} \sum_{k=1}^{D} Q_{j,k} \nonumber \\
    && \times \bigg( (C_x +C_c) Q_j \! + \! C_J \sum_{l=1}^{3} Q_{l} \bigg)  . \qquad \quad
\end{eqnarray}

We transform to the dimensionless coordinates $n = Q/2e$ and $\phi = 2\pi \Phi/ \Phi_0$ and perform the first quantization to obtain the following Hamiltonians
\begin{eqnarray}
 \hat  H_{\mathrm{ring}}  &=&  \frac{(2e)^2}{2} \mathbf{\hat n} \mathbb{C}^{-1} \mathbf{\hat n}  - E_J \sum_{j=1}^3 \cos  (\hat \phi_{j+1} - \hat \phi_j) , \qquad
 \label{eq:AppendHring0} \\
 \hat H_{\mathrm{wg}} &=& \sum_{j=1}^{3} \sum_{k=1}^{N}  E_{\mathcal C_{\mathrm{wg}}} \hat n_{j,k}^2 + E_{\ell_{\mathrm{wg}}} (\hat \phi_{j,k+1} - \hat \phi_{j,k})^2 , \qquad  \label{eq:AppendHwg0} \\
 \hat H_{\mathrm{int}} &=&  \frac{(2e)^2\mathcal C_c}{\mathcal C_{\mathrm{wg}} (C_x+C_c) C_{\Sigma}}  \sum_{j=1}^{3} \sum_{k=1}^{D} \hat n_{j,k} \nonumber \\
 && \times \bigg( (C_x+C_c) \hat n_j  +  C_J \sum_{l=1}^{3} \hat n_{l}  \bigg) , \qquad  \label{eq:AppendHint0}
\end{eqnarray}
where
$  E_{\mathcal C_{\mathrm{wg}}} = (2e)^2/(2 \mathcal C_{\mathrm{wg}}) $ and 
$ E_{\ell_{\mathrm{wg}}} = \Phi_0^2/(8 \pi^2 \ell_{\mathrm{wg}})$.

\subsection{Offset charges and external flux}
We include the offset charges and external flux to the circulator ring by simply making replacements in Eq.\ \eqref{eq:AppendHring0} as
$\mathbf{\hat n} \to (\mathbf{\hat n} - \mathbf{n}_x)$ with $\mathbf{n}_x = (n_{x_1}, n_{x_2}, n_{x_3})$ and $\hat\phi_{j+1}-\hat\phi_j \to \hat \phi_{j+1}-\hat \phi_j - \hat \phi_x/3 $, yielding
\begin{eqnarray}
\hat H_{\mathrm{ring}} &=& \frac{(2e)^2}{2} (\hat{\mathbf{n}} - \mathbf{n}_x) \mathbb{C}^{-1} (\hat{\mathbf{n}} - \mathbf{n}_x) \nonumber \\
&& - E_J \sum_{j=1}^{3} \cos(\hat \phi_j - \hat \phi_{j+1} - \tfrac{1}{3} \phi_x), \label{eq:AppendHringBias}
\end{eqnarray}
For the interaction Hamiltonian $\hat H_{\mathrm{int}}$ in Eq.\ \eqref{eq:AppendHint0}, we replace $\hat n_j \to \hat n_j - n_{x_j}$, so that  
\begin{eqnarray}
\hat H_{\mathrm{int}} &=&  \frac{(2e)^2\mathcal C_c}{\mathcal C_{\mathrm{wg}} (C_x+C_c) C_{\Sigma}}  \sum_{j=1}^{3} \sum_{k=1}^{D} \hat n_{j,k} \nonumber \\
&& \times \bigg( (C_x+C_c) (\hat n_j - n_{x_j} )  +  C_J \sum_{l=1}^{3} (\hat n_{l} - n_{x_l}) \bigg). \nonumber \\ \label{eq:AppendHintBais}
\end{eqnarray}

\subsection{Coordinate transformation} \label{subappend:CoordinateTrans}

Since the total number of Cooper pairs on the ring is conserved, we can reduce the number of dynamical variables of the circulator system from 3 to 2. To this end, we define new coordinates \cite{Koch10,Muller18}
\begin{equation}
   \mathbf{\hat n'} =  A \mathbf{\hat n}, \hspace{0.5cm} \pmb{\hat \phi}' = (A^T)^{-1} \pmb {\hat \phi}, \label{eq:AppendCoorTrans}
\end{equation}
where $\mathbf {\hat o} = (\hat o_1 \; \hat o_2\; \hat o_3)^{T}$ and
\begin{equation}
     A = \left( \begin{array}{ccc}
      1 & 0 & 0 \\
      0 &-1 & 0 \\
      1 & 1 & 1
    \end{array} \right).
\end{equation}
By this, $\hat H_{\mathrm{ring}}$ in Eq.\ \eqref{eq:AppendHringBias} in terms of new coordinates is
\begin{eqnarray}
   \hat H_{\mathrm{ring}} &=& E_{C\Sigma} \big( (\hat n'_1 - \tfrac{1}{2}(n_0 +n_{x_1}- n_{x_3}))^2 \nonumber  \\
&& + (\hat n'_2 + \tfrac{1}{2}(n_0 +n_{x_2}- n_{x_3}))^2 -\hat n'_1 \hat n'_2 \big) \nonumber \\
&& -E_J \big ( \cos(\hat \phi'_1 - \tfrac{1}{3}\phi_x) +\cos(\hat \phi'_2- \tfrac{1}{3}\phi_x) \nonumber \\
&& + \cos(\hat \phi'_1 + \hat \phi'_2  + \tfrac{1}{3}\phi_x) \big), \label{eq:AppendHring}
\end{eqnarray}
where $ n_0 = \langle \sum_{j=1}^3 \hat n_j \rangle_{\mathrm{gr}}$ is a conserved charge number evaluated from the expectation value of $\sum_{j=1}^3 \hat n_j$ in the ground state of $\hat H_{\mathrm{ring}}$ in Eq.\ \eqref{eq:AppendHring0} and $E_{C\Sigma}    =  (2e)^2/C_{\Sigma}.$
The expression of $\hat H_{\mathrm{int}}$ in Eq.\ \eqref{eq:AppendHintBais} is also altered to be
\begin{eqnarray}
   \hat H_{\mathrm{int}} = E_c \sum_{k=1}^{D} ( \hat n_{1,k} \hat q_1 + \hat n_{2,k} \hat q_2 + \hat n_{3,k} \hat q_3), \label{eq:Coupling}
\end{eqnarray}
where
\begin{eqnarray}
 &  \hat  q_1 = \hat n'_1 \!+\! n'_{x_1}, \hspace{0.2cm} \hat q_2 = - \hat n_2' + n'_{x_2}, \hspace{0.2cm} \hat q_3 = -\hat n'_1 + \hat n'_2 + n'_{x_3}, \nonumber \\
\label{eq:qCoupling} \\
  & \left\{  \begin{array}{ccl}
         n'_{x_1} &=& c_1 (n_0 - n_{x_2} - n_{x_3}) - c_2 n_{x_1} \\
         n'_{x_2} &=& c_1 (n_0 - n_{x_1} - n_{x_3}) - c_2 n_{x_2} \\
         n'_{x_3} &=& c_2 (n_0 - n_{x_3}) - c_1 (n_{x_1} + n_{x_2})
    \end{array} \right. , \label{eq:rescaledChargeBias}
\end{eqnarray}
with  $E_c = (2e)^2 \mathcal C_c/ (\mathcal C_{\mathrm{wg}}  C_{\Sigma})$, $c_1 = C_J/(C_x + C_c)$, and $c_2 = (C_J +C_x + C_c)/(C_x + C_c)$. We note that the above rescaled charge biases, because of the rotating wave approximation (used later) which considers only off-diagonal matrix elements of $\hat q_j$, are irrelevant in calculations.

\subsection{Waveguide normal modes and simplification of the interaction Hamiltonian}

Using the results in Ref. \cite{Le19}, we reexpress the waveguide Hamiltonian in Eq.\ \eqref{eq:AppendHwg0} in terms of its normal modes as
\begin{equation}
    \hat H_{\mathrm{wg}} = \sum_{j=1}^3 \sum_{k=0}^{\infty} \omega_{j,k} \hat a^\dag_{j,k} \hat a_{j,k}, 
\end{equation}
where $\omega_{j,k} = \omega_k = \pi k L^{-1} (\bar {\mathcal C}_{\mathrm{wg}} \bar{ \ell}_{\mathrm{wg}} )^{-1/2}$ with $L$, $\bar{\mathcal C}_{\mathrm{wg}}$, and $\bar{\mathcal \ell}_{\mathrm{wg}}$, respectively, the length, the capacitance per unit length, and the inductance per unit length of the waveguide. The interaction $\hat H_{\mathrm{int}}$ in the new waveguide modes is
\begin{eqnarray}
\hat H_{\mathrm{int}} &=& \sum_{j=1}^3 \sum_{k=1}^{\infty} g_k (\hat a^\dag_{j,k} + \hat a_{j,k}  ) \hat q_j, 
\end{eqnarray}
where $g_k = (2C_c/C_{\Sigma}) \sqrt{2 \omega_s/( R_K L \bar{\mathcal C}_{\mathrm{wg}})} $ with $R_K = h /(2e)^2 \approx 25.8 \mathrm{k}\Omega$ and $\hat q_j$ are given in Eq.\ \eqref{eq:qCoupling}. Taking the continuum limit \cite{Le19,Fan10} and expanding the lower limit of the frequencies to $-\infty$ \cite{GardinerBook04} the two above Hamiltonians are recast to
\begin{eqnarray}
\hat H_{\mathrm{wg}} &=& \sum_{j=1}^3 \int_{-\infty}^{\infty} d \omega\, \omega\, \hat a_j^\dag (\omega) \hat a_j (\omega), \label{eq:AppendHwg} \\
\hat H_{\mathrm{int}} &=& \sum_{j=1}^3 \int_{-\infty}^{\infty} d\omega g(\omega) (\hat a_j^\dag (\omega) + \hat a_j(\omega)) \hat q_j, \qquad \label{eq:AppendHint}
\end{eqnarray}
where $g(\omega) = ( 2C_c/ C_{\Sigma}) \sqrt{ 2\omega Z_{\mathrm{wg}}/  (\pi R_K)  }$ with $Z_{\mathrm{wg}} = \sqrt{\bar{\mathcal \ell}_{\mathrm{wg}}/ \bar{\mathcal C}_{\mathrm{wg}}}$ the waveguide impedance. We furthermore employ the rotating wave approximation and Markov approximation to simplify the interaction into
\begin{equation}
    \hat H_{\mathrm{int}} = \sum_{j=1}^3 \sqrt{\frac{\Gamma}{2\pi}} \int_{-\infty}^{\infty} d \omega (\hat a_j^\dag (\omega) \hat q_{j,-} + \hat a_j(\omega) \hat q_{j,+}  ),
\end{equation}
where $\sqrt{\Gamma/(2\pi)} = g(\omega_d)$ is the coupling strength evaluated at the driving frequency $\omega_d$.

\section{Adiabatic elimination} \label{append:AE}

In this Appendix, we present the adiabatic elimination procedure presented in Ref. \cite{Josh17} to find a new SLH triple defined in the slow subspace of the circulator ring.  
We denote the initial SLH triple of the system of interest with an upper bar $(\bar S, \bar L, \bar H)$. We define the operator 
\begin{equation}
    \bar K = - (i \bar H + \tfrac{1}{2} \textstyle \sum\nolimits_j \bar{L}_j^{\dag} \bar L_j  ).
\end{equation}
We decompose $\bar K$ as 
\begin{equation}
    \bar K = Y + A + B,
\end{equation}
where 
\begin{subequations}
\begin{eqnarray}
Y &=& P_1 \bar K P_1, \\
A &=& P_1 \bar K P_0 + P_0 \bar K P_1, \\
B &=& P_0 \bar K P_0,
\end{eqnarray}
\end{subequations}
with $P_0$ the projector onto the slow subspace and $P_1$ onto the fast subspace. We also decompose $\bar L_j$ as
\begin{equation}
    \bar L_j = F_j + G_j,
\end{equation}
where 
\begin{subequations}
\begin{eqnarray}
F_j &=& P_1 \bar L_j P_1 + P_0 \bar L_j P_1, \\
G_j &=& P_1 \bar L_j P_0 + P_0 \bar L_j P_0.
\end{eqnarray}
\end{subequations}
The operators (without an upper bar) in adiabatically eliminated subspace  are then given by
\begin{subequations}
\begin{eqnarray}
    K &=& -( i H + \tfrac{1}{2}\textstyle\sum\nolimits_j L_j^\dag L_j) = P_0 (B-A \tilde{Y} A)P_0 , \qquad \\
    L_j &=& (G_j - F_j \tilde{Y} A)P_0, \\
    H &=& i K + \tfrac{i}{2} \textstyle\sum\nolimits_j L_j^\dag L_j, \\
    S_{ij} &=& (F_i \tilde{Y} F_{\ell}^\dag + \delta_{i\ell}) \bar{S}_{\ell j} P_0, \label{eq:Sij}
\end{eqnarray}
\end{subequations}
where $\tilde{Y}$ satisfies
\begin{equation}
    Y \tilde Y = \tilde Y Y = P_1.
\end{equation}

For the circulator ring, its initial SLH triple is
\begin{eqnarray}
\bar H &=& \hat H_{\mathrm{ring}} + \hat H_{\mathrm{drive}}, \\
\bar L_j &=& \sqrt{\Gamma} \hat q^{(j)}_{-} + \beta_j  \mathbb{1}; \hspace{0.3cm} j=1,2,3, \\
\bar S &=& \mathrm{Diag} (\mathbb{1}, \mathbb{1}, \mathbb{1}), \label{eq:Sbar}
\end{eqnarray}
where $\hat H_{\mathrm{ring}}$ and $\hat H_{\mathrm{drive}}$ are expressed in a frame rotating with the driving frequency $\omega_d$: $\hat H_{\mathrm{ring}} = \sum_{k>0} (\omega_k - \omega_d ) \ket{k}\, \bra{k} $ and $\hat H_{\mathrm{drive}} = -\tfrac{i}{2} \sqrt{\Gamma} \sum_{j=1}^3 (\beta_j \hat q_{j,+} - \mathrm{H.c.} ) $. The ring slow-subspace and fast-subspace projectors are respectively defined as
\begin{eqnarray}
 P_0 &=& \ket{0}\bra{0}, \\
 P_1 &=& \sum_{k>0} \ket{k} \bra{k}.
\end{eqnarray}
Given the initial SLH triple and the projectors onto slow and fast subspaces, we carry out the computations outlined above to obtain the semi-analytical expression for the scattering matrix element $S_{ij}$ as given in \cref{eq:Selement}. Note that to simplify the operator $Y$, we make use of the results $|Q_{12}|, |Q_{21}|\ll |\gamma_1|, |\gamma_2|$ (see next Appendix) and $|\beta_j|^2 \ll \Gamma$ for weak coherent input fields.

\section{ decay rate comparison } \label{append:explainConditions}

\begin{figure}[h!]
    \centering
    \includegraphics[scale=0.95]{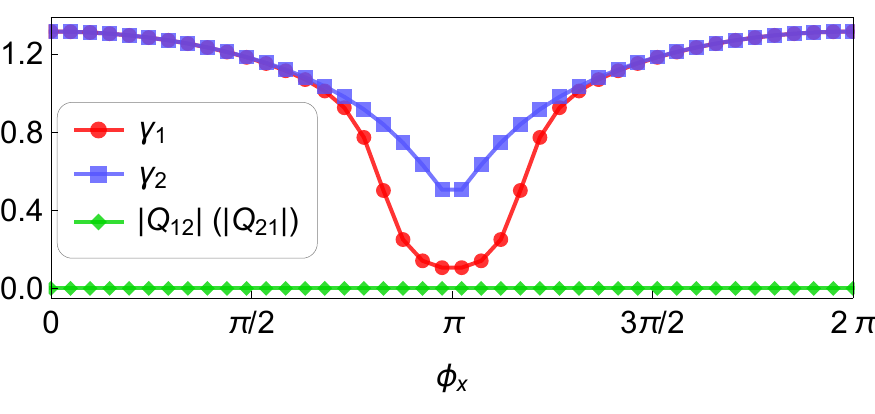}
    \caption{  $\gamma_1$, $\gamma_2$, and $|Q_{12}| \equiv |Q_{21}|$ as functions of the reduced external flux $\phi_x$, where $\gamma_1$ and $\gamma_2$ are respectively the decay rates of the excitations $\ket{1}$ and $\ket{2}$ and \mbox{$Q_{k \ell} = \sum_{j=1}^3 \langle 0 | \hat q_j | k \rangle \langle \ell | \hat q_j | 0 \rangle$}. Note that \mbox{$Q_{11} \equiv \gamma_1$} and \mbox{$Q_{22} \equiv \gamma_2$}. Relevant parameters are the same as in \cref{fig:circulator}b. }
    \label{fig:additionalInformation}
\end{figure}

In \cref{fig:additionalInformation} we compare $\gamma_1$, $\gamma_2$, and $|Q_{12}| \equiv |Q_{21}|$, where
\mbox{$Q_{k \ell} = \sum_{j=1}^3 \langle 0 | \hat q_j | k \rangle \langle \ell | \hat q_j | 0 \rangle$} 
for a symmetric circulator ring with identical charge biases of $1/3$. We clearly see that $|Q_{12}|  \simeq 0 $ while $ \gamma_1, \gamma_2>0$, thus justifying the approximation used to obtain the property in \cref{eq:ColumnSum}. Similar results hold for the case of an asymmetric circulator ring.

\section{Simulation parameters} \label{append:SimuPara}

\begin{table}[H]
    \centering
    \begin{tabular}{c|c}
    Parameter &  Value \\
    \hline
    $E_J/\hbar $ &  $ 2\pi \times 12.92\, \mathrm{GHz}$ \\
    \hline
     $C_J$    & $5.76\, \mathrm{fF}$  \\
     \hline
    $C_x$     & $5.95\, \mathrm{fF}$ \\
    \hline
    $C_c$ & $10.60 \, \mathrm{fF}$ \\
    \hline
    $E_{C_\Sigma}/\hbar $ & $2\pi \times \mathrm{4.58}\, \mathrm{GHz}$ \\
    \hline
    $Z_{\mathrm{wg}}$ (\cref{sec:Optimization}) & $50\, \Omega$ \\
    \hline 
    $Z_{\mathrm{wg}}$ (\cref{sec:QPPoisoning}) & $200\, \Omega$
    \end{tabular}
    \caption{Relevant parameters used for numerical simulations in this paper.}
    \label{tab:ParaVals}
\end{table}

\begin{figure}[h]
    \centering
    \includegraphics[width = 0.47 \textwidth]{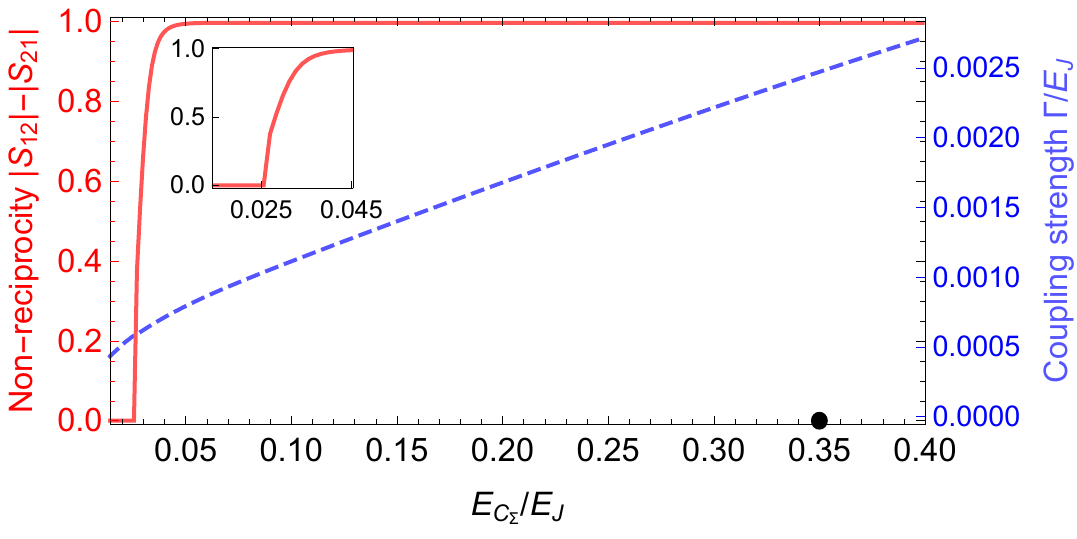}
    \caption{ Non-reciprocity represented by $|S_{12}| - |S_{21}|$ (solid red) and coupling strength $\Gamma$ (dashed blue) versus the ratio $E_{C_\Sigma}/E_J$. The inset shows the variation of $|S_{12}|-|S_{21}|$ when $E_{C_\Sigma}/E_J \leq 0.045 $; non-reciprocity disappears near $E_{C_\Sigma}/E_J = 0.025$. The coupling strength $\Gamma$ increases monotonically with $E_{C_\Sigma}/E_J$. The black dot indicates the value of $E_{C_\Sigma}/E_J$ ($=0.35$) used for simulations in the main text. The plot is generated for a symmetric ring. }
    \label{fig:CheckTransmonRegime1}
\end{figure}

Table \ref{tab:ParaVals} shows the values of the parameters used to perform numerical simulations in the present paper.  

In \cref{fig:CheckTransmonRegime1} we plot the non-reciprocity defined by $|S_{12}| - |S_{21}|$ (solid red) and the coupling strength $\Gamma$ (dashed blue) as functions of the ratio $E_{C_\Sigma}/E_J$ for a symmetric ring circulator. It is shown that non-reciprocity disappears at approximately $E_{C_\Sigma}/E_J = 0.025$ and $\Gamma$ is monotonically reduced when decreasing $E_{C_\Sigma}/E_J$.

\section{Variations of the working parameters along optimization} \label{append:Inspect}

\begin{figure}[ht]
    \centering
    \includegraphics[scale=0.82]{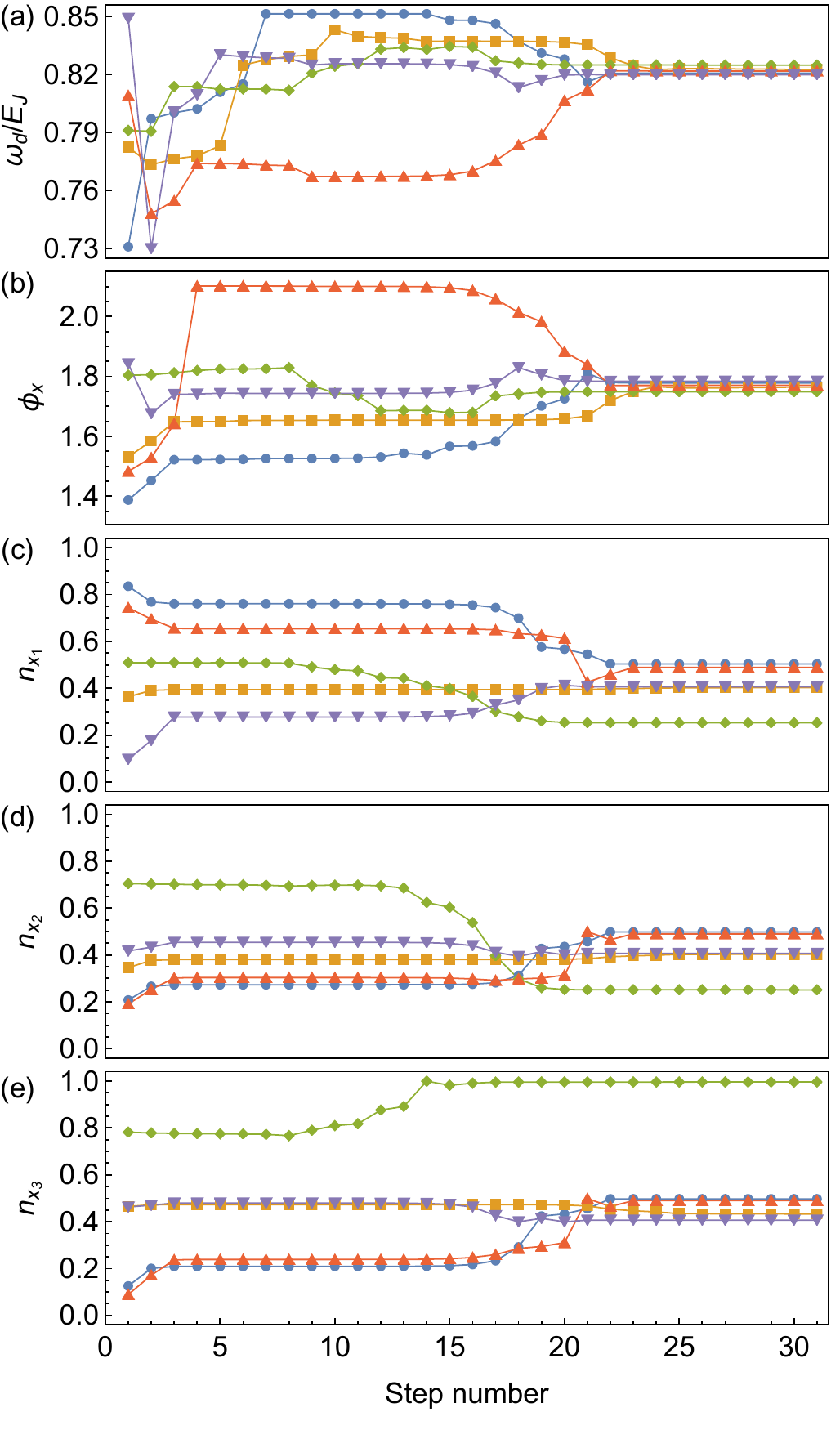}
    \caption{ Variations along the optimization  in \cref{subsec:SymNum} for a symmetric circulator ring of (a) the driving frequency $\omega_d$, (b) the external flux $\phi_x$, and (c) - (e) the three charge biases $n_{x_j}$. }
    \label{fig:OptParaSym}
\end{figure}

\begin{figure}[ht!]
    \centering
    \includegraphics[scale=0.82]{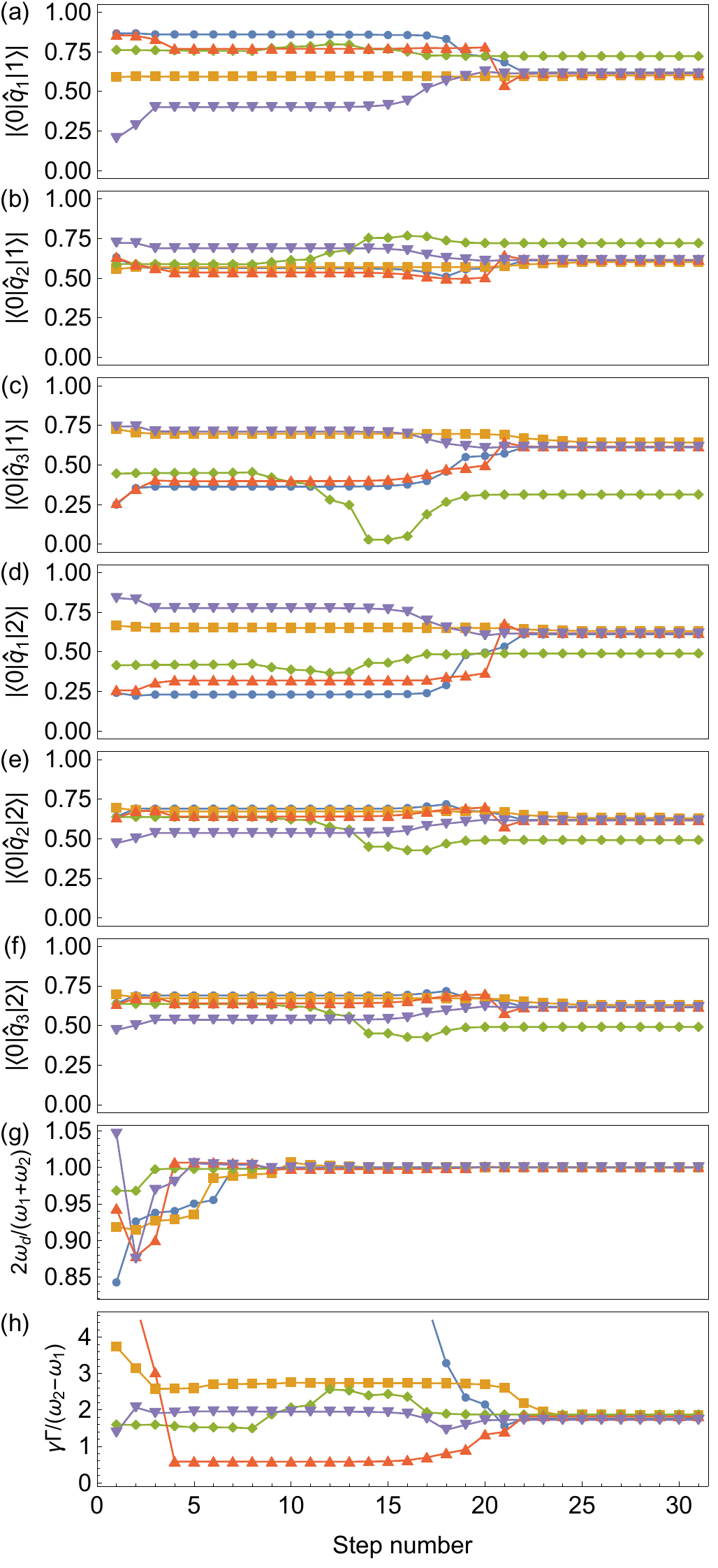}
    \caption{ Variations along the optimization  in \cref{subsec:SymNum} for a symmetric circulator ring of (a) - (f) the magnitudes of the coupling matrix elements $|\langle 0 | \hat q_j | k \rangle|$ for $j=1,2,3$ and $k=1,2$ to examine the condition in Eq. \eqref{eq:dipoleCondition}, (g)  the ratio $2\omega_d/(\omega_1 + \omega_2)$ to examine the condition in Eq.\
    \eqref{eq:driveCondition}, and (h) the ratio $\gamma \Gamma/(\omega_2 - \omega_1)$ to examine the condition in Eq.\ \eqref{eq:couplingCondition}. 
    }
   \label{fig:OptSymCheck}
\end{figure}

In \cref{fig:OptParaSym} we record the variations with respect to optimization steps of (a) the driving frequency $\omega_d$, (b) the external flux $\phi_x$, and (c) - (e) the three charge biases $n_{x_j}$ for a symmetric circulator ring after the optimization in \cref{subsec:SymNum}.
Meanwhile, in \cref{fig:OptSymCheck} we show the variations of (a) - (f) the magnitudes of the coupling matrix elements $| \langle 0 | \hat q_j |k \rangle|$ for $j=1,2,3$ and $k=1,2$  to inspect the condition in \cref{eq:dipoleCondition}, (g) the ratio $2 \omega_d/(\omega_1 + \omega_2)$ to inspect the condition in Eq.\  \eqref{eq:driveCondition}, and (h) the ratio $\gamma \Gamma/(\omega_2 - \omega_1)$ to inspect the condition in Eq.\ \eqref{eq:couplingCondition}.  

\begin{figure}
    \centering
    \includegraphics[scale=0.82]{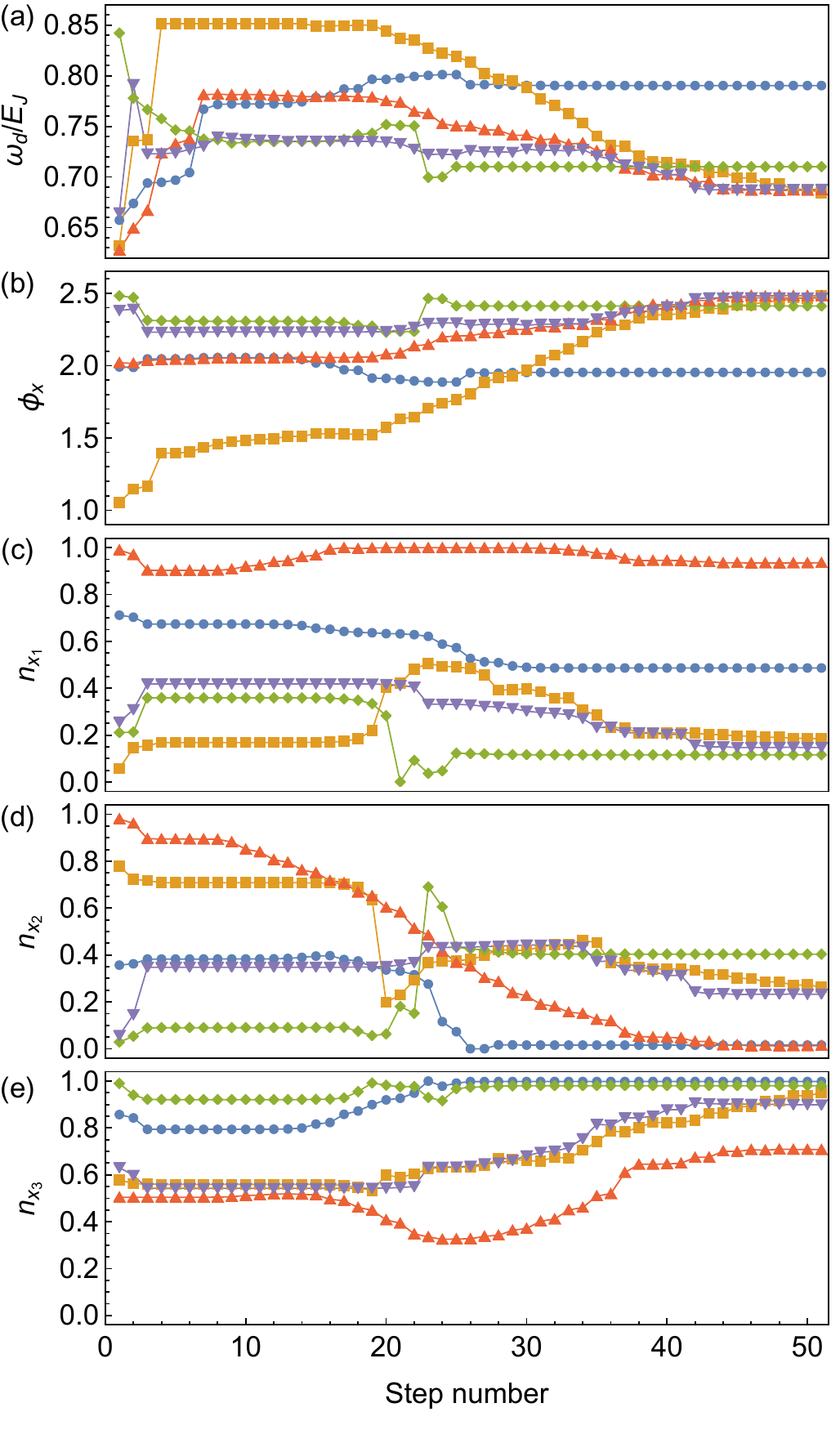}
    \caption{Similar to \cref{fig:OptParaSym} but for  an asymmetric circulator ring and the optimization in \cref{subsec:AsymNum}.}
    \label{fig:OptParaASym}
\end{figure}

\begin{figure}[t!]
    \centering
     \includegraphics[scale=0.82]{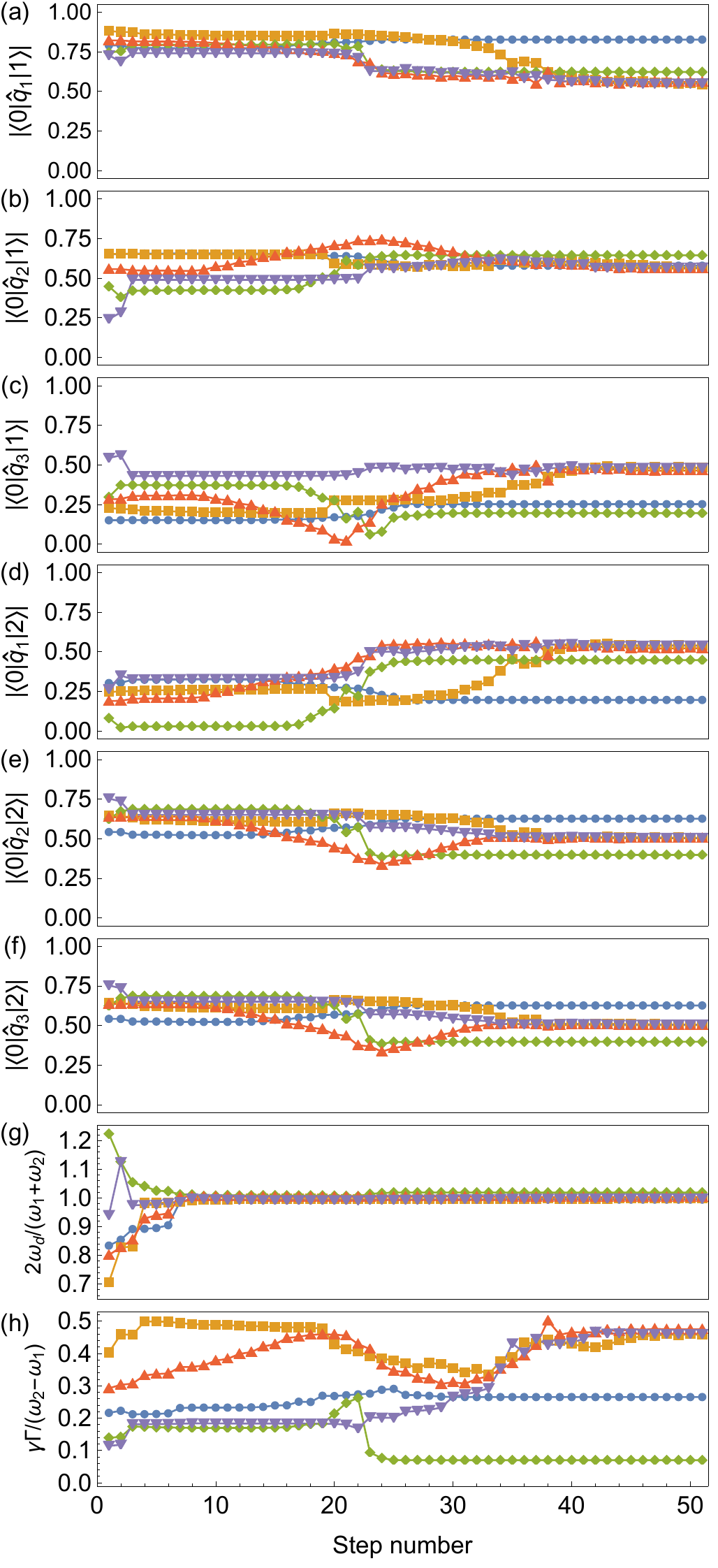}

    \caption{  Similar to \cref{fig:OptSymCheck} but for  an asymmetric circulator ring and the optimization in \cref{subsec:AsymNum}. }
   \label{fig:OptASymCheck}
   
   \end{figure}

Figures \ref{fig:OptParaASym} and \ref{fig:OptASymCheck} show the results for an asymmetric circulator ring.

\section{Derivation of the master equation in the presence of quasiparticle tunneling} \label{append:MEquasiparticle}

Here we derive the master equation for the ring density operator in the presence of both coupling to waveguides and coupling to quasiparticles as in Eq.\ \eqref{eq:MEqp}. The total waveguide-ring-quasiparticle Hamiltonian is \cite{Duy17}
\begin{equation}
    \hat H'_{\mathrm{tot}} = \hat H'_{\mathrm{ring}} + \hat H_{\mathrm{wg}} + \hat H_{\mathrm{int}} + \hat H_{\mathrm{qp}} + \hat H_{T}, \label{eq:AppendHtotQP}
\end{equation}
where $\hat H'_{\mathrm{ring}}$ is diagonal in sector blocks as in \cref{eq:RingHamiltonianBlock}, 
and $\hat H_{\mathrm{wg}}$ and $\hat H_{\mathrm{int}}$ are respectively given in Eqs.\ \eqref{eq:AppendHwg}, and \eqref{eq:AppendHint}. $\hat H_{\mathrm{qp}}$ is the  Hamiltonian of quasiparticles in the three ring islands \cite{Catelani11}
\begin{equation}
    \hat H_{\mathrm{qp}} = \sum_{j=1}^{3} \hat H^{(j)}_{\mathrm{qp}}, \hspace{0.5cm} \hat H^{(j)}_{\mathrm{qp}} = \sum_{n,\sigma} \epsilon_{n}^{(j)} \hat \alpha^{(j) \dag}_{n\sigma} \hat \alpha^{(j)}_{n\sigma}, 
\end{equation}
where $j$ indexes the islands, $\sigma =\, \uparrow, \downarrow $ denotes electron spins, $\epsilon^{(j)}_{n} = \sqrt{(\xi^{(j)}_n)^2 + (\Delta^{(j)})^2 }$ is the quasiparticle energy with 
$\xi^{(j)}_n$ the single-particle energy at level $n$ in the normal state of the island $j$ and $\Delta^{(j)}$ the gap parameter of that island, and $\hat \alpha_{n\sigma}^{(j)} $ ($\hat \alpha_{n\sigma}^{(j)\dag} $) is the quasiparticle annihilation (creation) operator.
$\hat H_T$ describes quasiparticle tunneling between the ring islands \cite{Catelani11,CatelaniPRB11}
\begin{eqnarray}
 \hat H_{\mathrm{T}} \! &=& \!  \sum_{j\ne j'} t_{jj'} \! \sum_{n,m,\sigma} \! \hat \alpha^{(j) \dag}_{n \sigma} \hat \alpha^{(j')}_{m\sigma} \nonumber \\
 && \! \times \big( e^{ \tfrac{i}{2} (\hat \varphi_j \!-\! \hat \varphi_{j'} ) } u^{(j)}_n u^{(j')}_m \!-\! e^{- \tfrac{i}{2} (\hat \varphi_j \!-\! \hat \varphi_{j'} ) } v^{(j')}_m v^{(j)}_n \big) + \mathrm{h.c}, \nonumber \\
\end{eqnarray}
where $t_{jj'} \ll 1$ is the tunneling amplitude between the islands $j$ and $j'$ and is determined via the junction conductance $g_{jj'} = 4\pi e^2 \nu^{(j)} \nu^{(j')} t^2_{jk}/\hbar $ with $\nu^{(j)}$ the density of states per spin direction in the island $j$, $\hat \varphi_j$ is the phase of the island $j$, and $u^{(j)}_n$ and $v^{(j)}_n$ are Bogoliubov amplitudes and are assumed to be real. 
In the low-energy limit \cite{Catelani11}, we approximate \mbox{$u^{(j)}_m \simeq v^{(j)}_n \simeq 1/\sqrt{2}$ }, so that 
\begin{equation}
    \hat H_{\mathrm{T}} =  \sum_{j\ne j'} t_{jj'} \sum_{n,m,\sigma}  \hat \alpha^{(j) \dag}_{n \sigma} \hat \alpha^{(j')}_{m\sigma} \hat T_{jj'} + \mathrm{h.c}, 
\end{equation}
where $\hat T_{jj'} = \sin ((\hat \varphi_j - \hat \varphi_{j'})/2) $ are the tunneling operators.
In terms of the new coordinates defined in Eq.\ \eqref{eq:AppendCoorTrans}, these operators are
\begin{equation}
    \hat T_{12} = \sin\big( \frac{\hat \phi'_1 + \hat \phi'_2}{2} \big), \hspace{0.25cm} \hat T_{23} = \sin \big( \frac{\hat \phi'_2}{2} \big), \hspace{0.25cm} \hat T_{31} = \sin \big( \frac{\hat \phi'_1}{2} \big).  
\end{equation}

We assume that the waveguides (the bosonic baths) are in vacuum \cite{Duy17} and consider the effect of weak coherent driving fields later. We also assume that quasiparticles in the ring islands are near equilibrium, so that
\begin{equation}
   \langle \hat \alpha^{(j)\dag}_{n\sigma} \rangle_{\mathrm{qp}} = \langle \hat \alpha^{(j)}_{n\sigma} \rangle_{\mathrm{qp}} = 0, \hspace{0.4cm}    \langle \hat \alpha^{(j)\dag}_{n\sigma} \hat \alpha^{(j')}_{n\sigma}  \rangle_{\mathrm{qp}} = \delta_{jj'} f^{(j)} [\epsilon_n^{(j)}],
\end{equation} 
where $f^{(j)} [\epsilon_n^{(j)}]$ is the distribution function of quasiparticles in the island $j$ which is assumed to be independent of spin. We decompose $\hat H'_{\mathrm{tot}}$ in \cref{eq:AppendHtotQP} into 
\begin{equation}
    \hat H'_{\mathrm{tot}} = \hat H_0 + \hat V,
\end{equation}
where  $\hat H_0 = \hat H'_{\mathrm{ring}} + \hat H_{\mathrm{wg}} + \hat H_{\mathrm{qp}}$ is the total unperturbed Hamiltonian and $\hat V = \hat H_{\mathrm{int}} + \hat H_T $ is a perturbation to $\hat H_0$.
Let $\varrho$ be the density operator of the total waveguide-ring-quasiparticle system and $\rho' = \mathrm{Tr}_{\mathrm{wg,qp}} (\varrho) $ be the ring density operator. In the interaction picture, the equation of motion of $\rho'$ is \cite{Duy17}
\begin{equation}
    \dot{\rho}'_I(t) = - \int_{-\infty}^t dt' \mathrm{Tr}_{\mathrm{wg,qp}} \big\{ [\hat V_I(t), [\hat V_I(t'), \varrho_I(t')]] \big\}, \label{eq:AppendMEInt}
\end{equation}
where
\begin{eqnarray}
 \hat V_I(t) \! &=& \! e^{- i \hat H_0 t} \hat V e^{i \hat H_0 t} \nonumber \\
 &=& \! \sum_{j=1}^3 \sqrt{\frac{\Gamma}{2\pi}} \int_{-\infty}^{\infty} d\omega \big( a_j^\dag(\omega) e^{i\omega t} \hat q_j^{-} (t)  \! + \! \mathrm{H.c.} \big) \nonumber \\
 && \! + \Big(\! \sum_{j\ne j'} \! t_{jj'} \!\! \sum_{n,m,\sigma} \! e^{i (\epsilon_n^{(j)} - \epsilon_m^{(j')})t} \hat \alpha^{(j)\dag}_{n\sigma} \hat \alpha^{j'}_{m\sigma} \hat T_{jj'}(t) \! + \! \mathrm{h.c} \! \Big). \nonumber \\
\end{eqnarray}
Here the operators $\hat q^-_{j}(t)$ and $\hat T_{jj'}(t)$ are 
\begin{eqnarray}
 \hat q_j^{-} (t) &=& \sum_s \sum_{k>k'} \langle k',s | \hat q_j | k,s \rangle e^{i \omega_{k,s;k',s}t} \ket{k',s}\!\bra{k,s}, \nonumber \\ \\
 \hat T_{jj'} (t) &=& \sum_{s, s'} \sum_{k,k'} \langle k',s' | \hat T_{jj'} | k,s\rangle e^{i \omega_{k,s;k',s'} t} \ket{k',s'}\!\bra{k,s}, \nonumber \\
\end{eqnarray}
where $s$ and $s'$ label the sector indices and $k$ and $k'$ label the ring eigenstate indices within each sector.
 
 The right-hand side of \cref{eq:AppendMEInt}, due to the double commutator, has four terms of which we consider only the term containing \mbox{$\hat V_I (t) \hat V_I (t') \varrho_I (t')$}. The other terms are evaluated similarly. We have
 \begin{eqnarray}
&&  - \int_{-\infty}^t dt' \mathrm{Tr}_{\mathrm{wg,qp}} \big\{ \hat V_I (t) \hat V_I (t') \varrho_I (t') \big\} \nonumber \\
&=&  - \frac{1}{2}  \sum_{j=1}^3 \sum_s \sum_{k > k'} \Gamma^{(j)}_{k,s;k',s} \ket{k,s}\!\bra{k',s} \ket{k',s}\!\bra{k,s} \rho'_{I}(t) \nonumber \\
&& - \frac{1}{2} \sum_{j\ne j'} \sum_{s,s'}\sum_{k,k'} \Gamma^{(jj')}_{k,s;k',s'}  \ket{k,s}\!\bra{k',s'} \ket{k',s'}\bra{k,s} \rho'_I (t), \nonumber \\
 \end{eqnarray}
where we have used the Born-Markov approximation and neglected fast oscillating terms, $\Gamma^{(j)}_{k,s;k',s}$ is the inner-sector relaxation rate
\begin{equation}
 \Gamma^{(j)}_{k,s;k',s}=  \Gamma |\langle k',s| \hat q_j |k,s \rangle|^2, 
\end{equation}
and $\Gamma^{(jj')}_{k,s;k',s'}$ is the inter-sector (i.e., quasiparticle-tunneling) rate
\begin{equation}
    \Gamma^{(jj')}_{k,s;k',s'} = | \langle k',s' | \hat T_{jj'} | k,s \rangle \big|^2 S_{\mathrm{qp}} (\omega_{k,s;k',s'}),
\end{equation}
with $\omega_{k,s;k',s'}$ the transition energy between the states $\ket{k,s}$ and $\ket{k',s'}$.
Here $S^{(jj')}_{\mathrm{qp}}(\omega)$ is the quasiparticle spectral density  and for $\omega>0$ is given by \cite{Catelani11}
\begin{eqnarray}
S^{(jj')}_{\mathrm{qp}} (\omega) &=& \frac{16 E^{(jj')}_J}{\pi} \int_{0}^{\infty} dx \frac{1}{\sqrt{x} \sqrt{x+\omega/\Delta}} \nonumber \\
&& \times \big( f^{(j)}[(1+x)\Delta] \{1-f^{(j')}[(1+x)\Delta + \omega] \} \big), \nonumber \\ \label{eq:AppendQPSpectral}
\end{eqnarray}
where $E_J^{(jj')}$ is the Josephson energy of the junction connecting the islands $j$ and $j'$. For $\omega<0$, in Eq.\ \eqref{eq:AppendQPSpectral} we make replacements $x \to x- \omega/\Delta$ and $\omega \to - \omega$. We consider equal populations on the islands, so $f^{(j)} = f^{(j')}$ simplifying $S_\mathrm{qp}^{(jj')}$ into Eq.\ \eqref{eq:qpSpectralDensity}. 

Using the above results, \cref{eq:AppendMEInt} is recast to
\begin{eqnarray}
\dot \rho'_I (t) &=& \sum_{j=1}\sum_{s}\sum_{k<k'} \Gamma^{(j)}_{k,s;k',s} \mathcal{D}[\ket{k',s}\!\bra{k,s}] \rho'_I (t) \nonumber \\
&& + \sum_{j\ne j'} \sum_{s,s'}\sum_{k,k'} \Gamma^{(jj')}_{k,s;k',s'} \mathcal{D}[\ket{k',s'}\!\bra{k,s}] \rho'_I(t). \qquad \quad
\end{eqnarray}
We transfer the master equation to the Schrodinger picture and add weak coherent driving fields \cite{Peropadre13,Breuer02}, yielding
\begin{eqnarray}
\dot \rho' (t) &=& -i [\hat H'_{\mathrm{ring}} -i \sqrt{\Gamma} \sum_{j} (\beta_j e^{-i \omega_{d} t } \hat q_{j,+} - \mathrm{H.c.} ), \rho'(t) ] \nonumber \\
    && + \sum_{j}\sum_{s}\sum_{k>k'} \Gamma^{(j)}_{k,s;k',s} \mathcal{D}[\ket{k',s}\!\bra{k,s}] \rho' (t) \nonumber \\
&& + \sum_{j\ne j'} \sum_{s,s'}\sum_{k,k'} \Gamma^{(jj')}_{k,s;k',s'} \mathcal{D}[\ket{k',s'}\!\bra{k,s}] \rho'(t), \qquad \quad \label{eq:AppendMEquasiparticleSchrodinger}
\end{eqnarray}
which is the master equation in Eq.\ \eqref{eq:MEqp}.

We unravel Eq.\ \eqref{eq:AppendMEquasiparticleSchrodinger} into a conditional master equation \cite{Stace03,GardinerBook04} 
\begin{eqnarray}
\dot \rho'_{c,0} (t) &=& -i [\hat H_{\mathrm{eff}}, \rho'_{c,0}(t) ] \nonumber \\
&& + \sum_{j}\sum_{s}\sum_{k>k'} \Gamma^{(j)}_{k,s;k',s} \mathcal{D}[\ket{k',s}\!\bra{k,s}] \rho'_{c,0} (t), \nonumber \\
\end{eqnarray}
where $\rho'_{c,0} (t)$ is the conditional system density operator conditioned on when no quasiparticle jumps happen, that is, it describes evolution within one quasiparticle sector in between incoherent jumps to the other sectors, and $\hat H_{\mathrm{eff}}$ is the effective no-jump Hamiltonian
\begin{eqnarray}
\hat H_{\mathrm{eff}} &=& \hat H'_{\mathrm{ring}} -i \sqrt{\Gamma} \sum_{j} (\beta_j e^{-i \omega_{d} t } \hat q_{j,+} - \mathrm{H.c.} ) \nonumber \\
&& - \frac{i}{2} \sum_{j\ne j'} \sum_{s,s'}\sum_{k,k'} \Gamma^{(jj')}_{k,s;k',s'}  \ket{k,s}\!\bra{k,s}. \label{eq:EffectiveJumpHamiltonian}
\end{eqnarray}
The non-Hermitian part of $\hat H_{\mathrm{eff}}$ (the second line in Eq.\ \eqref{eq:EffectiveJumpHamiltonian}) is the system self-damping \cite{GardinerBook04} given in terms of the sector-mixing operator \mbox{$\hat c_{k,s;k',s'} = \ket{k',s'}\bra{k,s}$} as $ \ket{k,s}\bra{k,s} = \hat c_{k,s;k',s'}^{\dag} \hat c_{k,s;k',s'}$. This self-damping term is
negligible compared to the Hermitian part of $\hat H_{\mathrm{eff}}$ (the first line in Eq.\ \eqref{eq:EffectiveJumpHamiltonian}), since $\Gamma^{(jj')}_{k,s;k',s'} \sim 1\, \text{kHz} $ (as estimated in \cref{subsec:QPFluctuations}) is much smaller than $ \omega_{k,s} \sim 2\pi \times  10\, \text{GHz}$ and $\Gamma \sim 2\pi \times 100\, \text{MHz}$ . 

\section{Sector fluctuations for an odd total charge-parity} \label{append:OddParity}

In this Appendix, we consider sector fluctuations when the total charge-parity of the ring islands is odd. The four quasiparticle sectors include \textsf{e-e(-o)}, \textsf{e-o(-e)}, \textsf{o-e(-e)}, and \textsf{o-o(-o)}. For a symmetric ring circuit, we find the fidelity $F(S_{\mathrm{sym}},S_{\mathrm{ideal}})$ for the sector \textsf{e-e} is optimized at $(\omega_d,\phi_x,n_{x_1},n_{x_2},n_{x_3}) = (0.77 E_J,2.11,1/3,1/3,5/6)$. We fix the three charge biases and plot the fidelity versus $\omega_d$ and $\phi_x$ for the four sectors. The results are almost identical to those in \cref{fig:SymSectorFluctuation}, so we do not show them here for brevity.

We repeat the above procedure for an asymmetric circuit with the same junction asymmetries as in \cref{fig:ASymSectorFluctuation} and observe that sector fluctuations are qualitatively analogous to the results in \cref{fig:ASymSectorFluctuation}.

\bibliography{circulatorPaper}

\end{document}